\def\arxiv{}
\ifdefined\arxiv
  \documentclass[11pt]{article}
  \usepackage[margin=1in]{geometry}
  \usepackage{xcolor}
  \usepackage{graphicx}
  \usepackage{amsthm}
  \usepackage{authblk}
  \usepackage{hyperref}
  \hypersetup{colorlinks=true,
    citecolor=blue!60!black,
    linkcolor=red!55!black,
    urlcolor=teal!70!black}
\else
  \documentclass{easychair}
\fi

\usepackage{amsmath}
\usepackage{amssymb}
\usepackage{multirow}
\usepackage{booktabs,subcaption,amsfonts}
\usepackage{xspace}
\usepackage{pgfplots}
\pgfplotsset{compat=1.18}
\usepackage{numprint}
\npthousandsep{,}
\usepackage{cancel}
\usepackage{rotating}
\usepackage{cleveref}

\newtheorem{theorem}{Theorem}

\newtheorem{definition}[theorem]{Definition}
\newtheorem{example}[theorem]{Example}

\newcommand{\set}[1]{\left\{
      \begin{array}{l}#1\end{array}
      \right\}}
\newcommand{\sset}[2]{\left\{~#1  \left|
      \begin{array}{l}#2\end{array}
    \right.     \right\}}
\newcommand{\etal}{\textit{et al.\ }}

\newcommand{\false}{\mathit{false}}

\newcommand{\true}{\mathit{true}}
\newcommand\tuple[1]{\langle #1 \rangle}

\newcommand{\pat}{\textsf{pat}}

\newcommand{\equivs}{\textit{equivs}}
\newcommand{\bee}{\textsf{BEE}}

\newcommand{\trivial}{{\scriptsize trivial}}

\title{Compact Partial Symmetry Breaking\\for Graph Search Problems}
\ifdefined\arxiv\else
\titlerunning{Compact Partial Symmetry Breaking}
\fi

\ifdefined\arxiv
\author[1]{Michael Codish}
\author[2]{Mikol{\'{a}}\v{s} Janota}
\author[3,4]{Peter J. Stuckey}
\affil[1]{Ben-Gurion University of the Negev, Israel}
\affil[2]{Czech Technical University, Czechia}
\affil[3]{Monash University, Australia}
\affil[4]{OPTIMA ARC Industrial Training and Transformation Centre, Australia}
\date{}
\else
\author{
  Michael Codish\inst{1}
\and
  Mikol{\'{a}}\v{s} Janota\inst{2}
\and
  Peter J. Stuckey\inst{3}$^,$\inst{4}
}
\institute{
  Ben-Gurion University of the Negev, Israel
\and
  Czech Technical University, Czechia
\and
  Monash University, Australia
\and 
OPTIMA ARC Industrial Training and Transformation Centre, Australia
}
\authorrunning{Codish, Janota, and Stuckey}
\fi

\begin{document}

\maketitle

\begin{abstract}\noindent
  Symmetry breaking in graph search problems remains challenging:
  complete symmetry breaking constraints are typically exponential in
  size, while partial approaches trade %
  precision for scalability.
  We introduce a strength-driven, redundancy-aware methodology for
  constructing compact partial symmetry breaking constraints based on
  graph patterns. Each pattern corresponds to a %
  Boolean clause over equality literals and can be integrated directly
  into CP or SAT models.
  Our approach incrementally selects strong, non-redundant graph
  patterns, yielding partial symmetry breaks that achieve substantial
  pruning of non-canonical graphs while remaining compact in
  size. This provides a principled framework for deriving lightweight
  partial symmetry breaking constraints which are significantly more
  precise than all published state-of-the-art alternatives.
  We evaluate the proposed constraints on graph instances with up to
  25 vertices using a redundancy ratio metric. The resulting partial
  symmetry breaks constitute only a small fraction of the size of
  complete symmetry breaking constraints, while delivering significant
  performance improvements across several graph search benchmarks
  compared to previously defined partial symmetry breaking methods.

\end{abstract}

\section{Introduction}

Graph search problems are about finding simple graphs with desired
structural properties. Such problems arise in many real-world
applications and are fundamental in graph theory.
Solving graph search problems is typically hard due to the enormous
search space and the large number of symmetries in graph
representation: Any graph obtained by permuting the vertices of a
solution (or a non-solution) is also a solution (or a non-solution),
which is isomorphic, or ``symmetric''.
To optimize the search we aim to restrict it to focus on one
``canonical'' graph from each isomorphism class.

One common approach to eliminate symmetries is to add \emph{symmetry
  breaking constraints} that are satisfied by at least one member of
each isomorphism class~\cite{crawford-kr96,Shlyakhter07,Walsh06}.
A symmetry breaking constraint is called \emph{complete} if it is
satisfied by exactly one member of each isomorphism class and
\emph{partial} otherwise.
In many cases, symmetry breaking constraints, complete or partial, are
expressed in terms of ``lex-constraints''. Each lex-constraint
corresponds to one symmetry, $\sigma$, which is a permutation on
vertices, and restricts the search space to consider assignments that
are lexicographically smaller than their permuted form obtained
according to $\sigma$.
If one considers the set of lex-constraints corresponding to all
permutations, then the corresponding symmetry break is complete but
too large to be of practical use.

Itzhakov and Codish~\cite{Itzhakov2016} observe that a complete
symmetry break for order-$n$ graphs can be defined in terms of a
number of lex-constraints which is considerably smaller than
$n$-factorial. They succeed in computing complete symmetry breaking
constraints of practical size for graphs with 10 or fewer
vertices. Dan\v{c}o~et~al.~\cite{danco2025} obtain similar results in
the context of finite models. However, this approach does not
scale. In subsequent work, Itzhakov~\etal\cite{CodishImps} introduce
the notion of ``lex-implications'', which refine lex-constraints, and
they compute complete symmetry breaks that are considerably smaller in
size. However, this approach still does not scale beyond graphs with
11 vertices.
It is known that breaking symmetry by adding lex-constraints to
eliminate symmetric solutions is intractable in
general~\cite{Babai1983,crawford-kr96}.  So, we do not expect to find
a complete symmetry break of polynomial size that identifies canonical
graphs that are lex-leaders.
Hence the interest in partial symmetry breaks.

Codish~\etal\cite{CodishMPS13,Codish2019} introduce a partial symmetry
break, which is equivalent to considering the quadratic number of
permutations that swap a pair of vertices
(transpositions). Rintanen~\etal\cite{RintanenR24} enhance this
approach for directed graphs.  This approach is widely applied and
turns out to work well in practice, despite eliminating only a small
portion of the symmetries.  However, when dealing with hard instances
of graph search problems, this constraint does not suffice.
Over the past decade, there has been little progress in the research
of partial symmetry breaking constraints for graph search
problems. Some attempts are made
in~\cite{DBLP:conf/cpaior/ItzhakovC23} and
in~\cite{DBLP:conf/cp/CodishGIS16}. However, there are no references in
the literature to applications which make use of the symmetry
breaks defined in these papers.

In a recent paper, Codish and Janota~\cite{CodishJanotaCP25}
introduce the notion of a graph pattern for symmetry breaking.
Graph patterns are simple atoms which provide a concise representation
for (large) sets of non-canonical graphs.
This also facilitates a set-covering perspective on symmetry breaking
as described in~\cite{cpaior2025}. A complete symmetry break should
cover all of the non-canonical graphs.  A partial symmetry break
should cover ``most'' of the non-canonical graphs.

This paper is about scalable and precise partial symmetry breaks for
graphs constructed using graph patterns. The idea is to repeatedly
select graph patterns based on two notions: strength of patterns, and
non-redundancy of patterns with respect to those already selected.
The proposed construction results in symmetry breaks which are
scalable both in terms of the cost to compute them as well as in terms
of their size. Precision is measured in terms of redundancy
ratio~\cite{Heule2019}, $\rho$, that is the ratio between the number
of graphs satisfying the symmetry break, and the number of canonical
graphs. This means that, for example, when enumerating the solutions
of a graph search problem with a symmetry breaking predicate with
redundancy ratio $\rho$, we can expect to encounter on average $\rho$
solutions for every canonical solution.

Applying the techniques proposed in~\cite{CodishMPS13,Codish2019} and
in~\cite{CodishImps}, one can obtain complete symmetry breaks for graphs of
order up to $n=10$ and $n=11$, respectively.  For $n=10$, the approach
of~\cite{CodishImps} results in a complete symmetry break which
consists of $25{,}263$ graph patterns which is basically the same
number of CNF clauses.
Using involutions as reported in~\cite{CodishJanotaCP25}, one can
obtain a partial symmetry break for order-10 graphs with $12{,}616$
graph patterns and a redundancy ratio of $1.04$. But this approach
also does not scale.
In this paper we show a scalable symmetry break consisting of
$2{,}273$ graph patterns for $n=10$ with a redundancy ratio,
$\rho=1.34$. That means on average, for every canonical graph we get
$1.34$ graphs instead of $1$. Moreover, we demonstrate the feasibility
of our approach for graphs of sizes up to 25.

\section{Preliminaries and Notation}\label{sec:prelim}

\noindent\textbf{Representing Graphs:~~}
Throughout this paper we consider simple graphs, i.e.\ undirected
graphs with no self-loops.
The adjacency matrix of a graph $G$ is an $n\times n$ Boolean
matrix. The element at row $i$ and column $j$ is $\true$ if and only
if $(i,j)$ is an edge. The list of edge variables of a graph $G$ is
denoted $edges(G)$. It consists of the elements in the upper (or
lower) triangle of the matrix representation in some fixed order.
Throughout this paper we assume that the list consists of the
$\binom{n}{2}$ elements obtained as the concatenation of the
\emph{columns} of the upper triangle of $G$.
We will clarify the motivation for the column-wise ordering later in
this section. 
In abuse of notation, we let $G$ denote a graph in
any of its representations.
An \emph{unknown graph} of order-$n$ is represented as an $n\times n$
adjacency matrix of Boolean variables which is symmetric and has the
values $\false$ (denoted by 0) on the diagonal, or as the
corresponding list of edge variables.
The following example specifies two order-$4$ unknown graphs as
adjacency matrices and as lists of edge variables.

\begin{example}\label{example:graph-representation}
  The following depicts two  unknown, order-4, graphs $G_1$ and $G_2$ and their
  representations as lists of edges.

  \medskip\noindent
  \resizebox{\textwidth}{!}{%
  {\small\begin{tabular}{p{.253\textwidth}p{.285\textwidth}l}
 {$\mathbf{G_1=}\left[\begin{matrix}
    0   & x_1 & x_2 & x_4 \\
    x_1 & 0   & x_3 & x_5 \\
    x_2 & x_3 & 0   & x_6 \\
    x_4 & x_5 & x_6 & 0 
\end{matrix}\right]$}
&
  {$\mathbf{G_2=}\left[\begin{matrix}
    0 & x_1 & x_4 & x_2 \\
    x_1 & 0 & x_5 & x_3 \\
    x_4 & x_5 & 0 & x_6 \\
    x_2 & x_3 & x_6 & 0 
\end{matrix}\right]$}
&
 $\begin{array}{l}
   \text{edges}(G_1)=\tuple{x_1,x_2,x_3,x_4,x_5,x_6}\\  
   \text{edges}(G_2)=\tuple{x_1,x_4,x_5,x_2,x_3,x_6}
\end{array}$
\end{tabular}
}}
\end{example}

\medskip\noindent\textbf{Ordering Graphs:~~}
Let $G_1,G_2$ be known or unknown graphs with $n$ vertices. Then,
$G_1 \leq G_2$ if and only if $edges(G_1)\leq_{lex} edges(G_2)$ where
$\leq_{lex}$ denotes the standard lexicographic ordering.
When $G_1$ and $G_2$ are unknown graphs, then the lexicographic
ordering, $G_1\leq G_2$, specifies a \emph{lexicographic order
  constraint} over the variables in $G_1$ and $G_2$.

\begin{example}\label{example:lex-constraint}
  Consider the two graphs, $G_1$ and $G_2$ from
  Example~\ref{example:graph-representation}.
  The lexicographic order constraint $G_1\leq G_2$ is 
  $\tuple{x_1,x_2,x_3,x_4,x_5,x_6} \leq_{lex}
    \tuple{x_1,x_4,x_5,x_2,x_3,x_6}$
  which can be simplified as described by
  Frisch~\etal\cite{Frisch03} to:
  $\tuple{x_2,x_3} \leq_{lex} \tuple{x_4,x_5}$.
\end{example}

For Boolean strings $\bar a=\tuple{a_1,\ldots, a_m}$ and
$\bar b=\tuple{b_1, \ldots, b_m}$ and $1\leq i\leq m$, we denote
  \begin{equation}
    \label{eq:smaller-i}
    \bar a<_{lex}^i \bar b \Leftrightarrow
    \set{(a_1=b_1),\ldots, (a_{i-1}=b_{i-1}),
      (a_i=0), (b_i=1)}
  \end{equation}
  We sometimes refer to  $\bar a<_{lex}^i \bar b$ as the set of
  equations on the right side of Equation~\eqref{eq:smaller-i}.
  We denote $G_1<^i G_2$ if $edges(G_1)<_{lex}^i edges(G_2)$.

  \begin{example}\label{example:lex-implication}
  Consider the two graphs, $G_1$ and $G_2$ from
  Example~\ref{example:graph-representation}. Then,
  \[ G_1<^3 G_2 \Leftrightarrow \set{ (x_1{=}x_1), (x_2{=}x_4),
    (x_3{=}0), (x_5{=}1)}
  \]
  \[ G_1<^5 G_2 \Leftrightarrow \set{ (x_1{=}x_1), (x_2{=}x_4),
    (x_3{=}x_5), (x_4{=}x_2), (x_5{=}0), (x_3{=}1) }
  \]
  Observe that the second constraint ($G_1<^5 G_2$) is not
  satisfiable. 
  \end{example}

\medskip\noindent\textbf{Permutations:~~}
The group of permutations on $\{1 \ldots n\}$ is denoted $S_n$.
We represent a permutation $\pi \in S_n$ as a sequence of length $n$
where the i$^{th}$ element indicates the value of $\pi(i)$.  For
example: the permutation $[2,3,1] \in S_3$ maps as follows:
$\set{1 \mapsto 2, 2 \mapsto 3, 3 \mapsto 1}$.

A \emph{transposition} is a permutation that swaps two elements and
leaves all other in place. For example, $[4,2,3,1]$ is the
transposition that swaps elements $1$ and $4$. 
An \emph{involution} is a permutation $\pi$ such that $\pi\circ\pi$ is
the identity. Involutions are permutations that swap any set of pairs
of disjoint elements. For example, $[4,3,2,1]$ is the involution which
swaps the pairs $\set{1,4}$ and $\set{2,3}$.

Permutations act on graphs and on unknown graphs in the natural
way. For a graph and also for an unknown graph $G$,
viewing $G$ as an adjacency matrix, given a permutation $\pi\in S_n$,
then $\pi(G)$ is the adjacency matrix obtained by mapping each element
at position $(i,j)$ to position $(\pi(i),\pi(j))$ (for $1\leq i,j\leq
n$). 
Two graphs $G,H$ are \emph{isomorphic} if there exists a
permutation $\pi \in S_n$ such that $G=\pi(H)$.

Let $G$ be an order-$n$ graph and $\pi_v$ a permutation on the
vertices of $G$. Let $edges(G)=\tuple{x_1,\ldots,x_m}$ and 
$edges(\pi(G))= \tuple{y_1,\ldots,y_m}$ (for $m= {{n}\choose{2}}$).
The permutation $\pi_e$ such that
$\pi_e(\tuple{x_1,\ldots,x_m})=\tuple{y_1,\ldots,y_m}$ is called the
edge permutation corresponding to $\pi_v$.
Often we write $\pi_v$ (or $\pi_e$) to indicate that a
permutation applies on the vertices (or edges) of a graph.

\begin{example}\label{example:graph-permutation}
  Consider the two graphs, $G_1$ and $G_2$ from
  Example~\ref{example:graph-representation} and consider the
  permutation $\pi_v = [1,2,4,3]$. Then $G_2=\pi(G_1)$. Recall that 
 $\text{edges}(G_1)=\tuple{x_1,x_2,x_3,x_4,x_5,x_6}$ and
 $\text{edges}(G_2)=\tuple{x_1,x_4,x_5,x_2,x_3,x_6}$. So,
 $\pi_e=[1,4,5,2,3,6]$ is the edge permutation corresponding to $\pi_v$.

\end{example}

\medskip\noindent\textbf{Symmetry Breaks:~~}
A \emph{symmetry break} for graph search problems is a predicate, $\psi(G)$,
on a graph $G$, which is satisfied by at least one graph in each
isomorphism class of graphs.
If $\psi$ is satisfied by exactly one graph in each isomorphism class,
then we say that $\psi$ is a \emph{complete symmetry
  break}. Otherwise, it is \emph{partial}.
In our setting, the canonical graphs, are the minimal (lexleader)
graphs of the isomorphism classes of graphs. 
In this setting, when lexicographic order is defined in terms of the
column-wise concatenation of edge variables, canonical graphs of order
$n$ can be extended to canonical graphs of order
$n+1$~\cite{ItzhakovC20}. This facilitates an incremental approach
when generating symmetry breaking constraints.

Heule~\cite{Heule2019} defines the notion of \emph{redundancy ratio},
which we denote $\rho(\psi)$, to measure the precision of $\psi$.
This is the ratio between the number of graphs that satisfy $\psi(G)$
and the number of isomorphism classes.  One can view $\rho(\psi)$ as
the average number of graphs per isomorphism class that are not
eliminated by $\psi$.

Table~\ref{table:state-of-the-art} illustrates (some of) the current
``state of the art'' for partial and complete lex-leader static
symmetry breaks. All of the results presented in this table are
obtained within a 24 hour cutoff. The missing entries are timeouts.
On the left, the columns titled \texttt{trans} are about symmetry
breaks based on transpositions as presented in
\cite{CodishMPS13,Codish2019}. These are the most widely applied
partial symmetry breaks because (1) they are scalable (polynomial in
size), and (2) they are easy to implement by imposing a
lexicographic order on the rows (or columns) of the adjacency
matrix. %
The problem with this symmetry break is that it is very imprecise.

On the right, the column titled \texttt{comp} is about the complete
symmetry breaks described in~\cite{CodishImps}. These are complete, so
their redundancy ratio is $1.00$. We have crossed out the entry for
$n=11$. This result is given in~\cite{CodishImps} where it is reported
to take 8.44 days to compute.

In the middle, the columns titled \texttt{con inv}, \texttt{disj
  inv}, and ~\texttt{inv} are about symmetry breaks based on
graph patterns derived from specific types of permutations:
consecutive, disjoint involutions and involutions, as defined in
\cite{CodishJanotaCP25}. 
For each of the symmetry breaks, the columns titled ``size'' report
the number of lex constraints of the form $G<^i\pi(G)$. The columns
titled ``$\rho$'' detail the redundancy ratio of the corresponding
symmetry break. For smaller values of $n$, $\rho$ is computed using an
exact model counter \texttt{ganak-2.5.2}~\cite{ganak}, and for larger
values of $n$, $\rho$ is computed using an approximate model counter
\texttt{approx-4.2}~\cite{approxmodelcounter}. In the latter case, the
value is annotated in the table by the symbol ``$\approx$''.

Table~\ref{table:state-of-the-art} illustrates the challenge for
computing (partial) symmetry breaks.  
Complete symmetry breaks quickly become impractical, while the
simplest partial break \texttt{trans} quickly leads to high redundancy
ratios.
While involutions can generate very high quality partial
symmetry breaks, they also quickly become challenging to compute.  The
strongest form \texttt{inv} hardly scales beyond the complete break;
the disjunctive form \texttt{disj inv} scales a little further still
with low redundancy; and the weakest form \texttt{con inv} scales
better but generates high redundancy ratios as $n$ grows.

\begin{table}
  \caption{State-of-the-art symmetry breaks compared in terms of size
    (number of lex constraints $G <^i \pi(G)$) and redundancy ratio (sometimes approximately computed).}
  \label{table:state-of-the-art}  \centering
\begin{tabular}{||l||r|r||r|r||r|r||r|r||r||}
  \hline
  n &
    \multicolumn{2}{|c||}{~~\texttt{trans\,\cite{Codish2019}}} &
    \multicolumn{2}{|c||}{~~\texttt{con inv\,\cite{CodishJanotaCP25}}} &
    \multicolumn{2}{|c||}{~~\texttt{disj inv\,\cite{CodishJanotaCP25}}} &
    \multicolumn{2}{|c||}{~~\texttt{inv\,\cite{CodishJanotaCP25}}} &
    \multicolumn{1}{|c||}{\texttt{comp\,\cite{CodishImps}}} \\
    \hline
  &
     size & $\rho$ &
     size & $\rho$ &
     size & $\rho$ &
     size & $\rho$ &
     size  \\
  \hline
  4  &   6 & $1.00$            &      6 & $1.00$          &    6  & $1.00$         & 6     & $1.00$          &      6 \\
  5  &  13 & $1.26$            &     13 & $1.06$          &    15 & $1.06$         & 16    & $1.00$          &     17 \\
  6  &  24 & $1.77$            &     29 & $1.16$          &    32 & $1.12$         & 39    & $1.00$          &     42 \\
  7  &  40 & $3.02$            &     64 & $1.46$          &    72 & $1.31$         & 113   & $1.01$          &    127 \\
  8  &  62 & $5.39$            &    134 & $1.95$          &   166 & $1.53$         & 399   & $1.02$          &    473 \\
  9  &  91 & $9.42$            &    269 & $2.76$          &   399 & $1.83$         & 2014  & $1.03$          &   2826 \\
  10 & 128 & $15.34$           &    526 & $3.97$          &   996 & $2.20$         & 12616 & $1.04$          &  25263 \\
  11 & 174 & $23.52$           &   1005 & $5.71$          &  2499 & $2.65$         & 81439 & ${\approx}1.12$  &  \cancel{274109}   \\
  12 & 230 & ${\approx}32.47$   &   1891 & ${\approx}7.39$  &  6244 & ${\approx}3.02$ &       &       &        \\
  13 & 297 & ${\approx}48.77$   &   3511 & ${\approx}12.19$ & 15447 & ${\approx}4.18$ &       &       &        \\
  14 & 376 & ${\approx}69.44$   &   6442 & ${\approx}16.74$ & 37721 & ${\approx}4.26$ &       &       &        \\
  15 & 468 & ${\approx}71.61$   &  11693 & ${\approx}19.96$ & 90876 & ${\approx}6.16$ &       &       &        \\
  16 & 574 & ${\approx}106.25$  &  21028 & ${\approx}28.33$ &       &      &       &       &        \\
  17 & 695 & ${\approx}191.28$  &  37509 & ${\approx}43.41$ &       &      &       &       &        \\
  18 & 832 & ${\approx}221.25$  &  66427 & ${\approx}49.64$ &       &      &       &       &        \\
  19 & 986 & ${\approx}256.27$  & 116888 & ${\approx}56.48$ &       &      &       &       &        \\
  20 & 1158& ${\approx}345.82$  &        &                 &       &      &       &       &        \\
  \hline
\end{tabular}

\end{table}

\section{Graph Patterns}

Graph patterns were introduced in~\cite{cpaior2025} and are similar to
the notion of lex-implications introduced in~\cite{CodishImps}. 
A \emph{graph pattern} is a partially instantiated graph $G$ (some
elements are variables) such that all instances of $G$ are
non-canonical. We typically represent a graph pattern $G$ using the
list notation $\text{edges}(G)$.
Formally, graph patterns derive from permutations as stated in the
following definition which is adapted from~\cite{cpaior2025}.

\begin{definition}[graph patterns~\cite{cpaior2025}]\label{def:patterns}
  Let $\pi$ be a permutation, $G$ be an unknown graph of order-$n$
  with $edges(G)=\tuple{x_1,\ldots,x_m}$,
  $edges(\pi(G))= \tuple{y_1,\ldots,y_m}$, and let $1\leq i\leq m$.
  The graph pattern, $\pat_i(\pi)$, is the  (most
  general) solution of equations from the right side in
  Equation~\eqref{eq:smaller-i} for the case of
  $edges(\pi(G))<_{lex}^i edges(G)$. We will refer to these equations
  as the ``graph pattern equations''.
  If the equations have no solution, then we denote
  $\pat_i(\pi)=\bot$.
\end{definition}

The graph pattern $p=\pat_i(\pi)$ represents the set of
(non-canonical) graphs $G$ that get smaller at position $i$ with
$\pi$.
Namely, graphs $G$ such that $\pi(G)<^iG$.
If $p$ represents $G$ we say that $p$ covers $G$.

\begin{example}
  Consider the two graphs, $G_1$ and $G_2$ from
  Example~\ref{example:graph-representation}. As stated in
  Example~\ref{example:graph-permutation}, for $\pi = [1,2,4,3]$,
  $G_2=\pi(G_1)$ and as demonstrated in
  Example~\ref{example:lex-implication},
  \[ G_1<^3 G_2 \Leftrightarrow \set{ (x_1{=}x_1), (x_2{=}x_4),
      (x_3{=}0), (x_5{=}1)}.
  \]
  Definition~\ref{def:patterns} sets
  $\pat_3(\pi) = [x_1,x_2,0,x_2,1,x_6]$ as determined by the (most
  general) solution of the above graph pattern equations.
\end{example}

Note that $\pat_i(\pi)$ is always a legal graph pattern (represents
only non-canonical graphs) because graphs that get smaller under some
permutation are trivially non-canonical.
Table~\ref{table:strength} details, a set of 18 graph patterns for
order-5 graphs. These 18 graph patterns form a complete symmetry
break for order-5 graphs. Namely, their instances are exactly all of
the corresponding non-canonical graphs.
The first column specifies a pair consisting of a vertex permutation
$\pi_v$ and an index $i$. The second column details the corresponding
graph pattern $\pat_i(\pi_v)$.

 \begin{table}
   \caption{Graph patterns for order-5 graphs with corresponding strengths and equations.}
   \label{table:strength}
   \centering
   
 \begin{tabular}{cccl}
   \hline
   $\pi_v-i$  &  $p=\pat_i(\pi_v)$ & $str(p)$ & $\pi_e$ (prefix) \\
   \hline
    $[3,2,1,4,5]-1$ & $[1,A,0,B,C,D,E,F,G,H]$ & $8$ & $[~]$ \\
    $[3,1,2,4,5]-1$ & $[1,0,A,B,C,D,E,F,G,H]$ & $8$ & $[~]$ \\
    $[1,2,4,3,5]-2$ & $[A,1,B,0,C,D,E,F,G,H]$ & $8$ & $[1]$ \\
    $[1,2,3,5,4]-4$ & $[A,B,C,1,D,E,0,F,G,H]$ & $8$ & $[1,2,3]$ \\
   \hline
    $[3,5,4,2,1]-2$ & $[A,1,B,C,D,0,E,F,A,G]$ & $7$ & $[9]$ \\
    $[2,3,1,4,5]-2$ & $[0,1,0,A,B,C,D,E,F,G]$ & $7$ & $[3]$ \\
    $[2,1,3,4,5]-4$ & $[A,B,B,1,0,C,D,E,F,G]$ & $7$ & $[1,3,2]$ \\
    $[1,2,3,5,4]-5$ & $[A,B,C,D,1,E,D,0,F,G]$ & $7$ & $[1,2,3,7]$ \\
   \hline
    $[4,3,1,2,5]-3$ & $[A,0,1,0,B,A,C,D,E,F]$ & $6$ & $[6,4]$ \\
    $[2,4,1,3,5]-3$ & $[A,A,1,0,A,B,C,D,E,F]$ & $6$ & $[5,1]$ \\
    $[3,1,2,4,5]-4$ & $[A,A,A,1,B,0,C,D,E,F]$ & $6$ & $[2,3,1]$ \\
    $[2,1,4,3,5]-7$ & $[A,B,C,C,B,D,1,0,E,F]$ & $6$ & $[1,5,4,3,2,6]$ \\
    $[1,3,2,4,5]-8$ & $[A,A,B,C,D,D,E,1,0,F]$ & $6$ & $[2,1,3,4,6,5,7]$ \\
    $[1,2,4,3,5]-9$ & $[A,B,C,B,C,D,E,F,1,0]$ & $6$ & $[1,4,5,2,3,6,7,8]$ \\
   \hline
  $[3,1,4,2,5]-5$ & $[0,0,A,A,1,0,B,C,D,E]$ & $5$ & $[2,6,4,3]$ \\
    $[2,1,4,3,5]-9$ & $[A,B,C,C,B,D,E,E,1,0]$ & $5$ & $[1,5,4,3,2,6,8,7]$ \\
   \hline
    $[2,3,5,1,4]-5$ & $[A,0,A,A,1,B,C,0,A,D]$ & $4$ & $[3,8,9,1]$ \\
    $[3,5,1,2,4]-6$ & $[0,A,B,B,C,1,B,C,0,D]$ & $4$ & $[9,2,7,3,8]$ \\
   \hline
 \end{tabular}
\end{table}

The strength of a graph pattern $p=\pat_i(\pi_v)$, denoted $str(p)$, is
a measure of the number of (non-canonical) graphs that it
represents. More precisely, it is the \texttt{log} of this number, or
equivalently, it is the number of (distinct) variables in $p$. The
strongest graph patterns are of strength $\binom{n}{2}-2$.
The third column of Table~\ref{table:strength} details the
strength of the corresponding  graph patterns which is precisely the
number of distinct variables in each $p=\pat_i(\pi_v)$. Every instance of $p$ is
a non-canonical graph. The rows of Table~\ref{table:strength} are
partitioned to highlight different patterns of the same strength.
It is always the case that $str(p)\leq \binom{n}{2}-2$. This is
because, by Definition~\ref{def:patterns}, $p$ will always contain at
least one ``1'' and at least one ``0''.

The fourth column of Table~\ref{table:strength}, titled $\pi_e$ (prefix),
specifies the length $i-1$ prefix of the edge permutation $\pi_e$
corresponding to the vertex permutation $\pi_v$. This prefix determines
a set of graph pattern equations. For example, read the prefix
$[9,2,7,3,8]$ from the last row in the table as the set of equations:
$eqs(p)=\{x_1{=}x_9, x_2{=}x_2, x_3{=}x_7, x_4{=}x_3, x_5{=}x_8\}$
(without the two equations assigning the $i^{th}$ edges from $G$ and
$\pi(G)$ to 1 and 0).
These equations determine which positions in $p$ are equal.

More formally, let $\pi_v-i$ specifiy a graph pattern $p$ and let
$\pi_e$ be the edge permutation corresponding to $\pi_v$. Then the set
of graph pattern equations which determine $p$ is
$eqs(p)=\sset{x_j=x_{\pi_e(j)}}{1\leq j<i}$.

\smallskip
\noindent
\textbf{Question:} What weakens a graph pattern $p=\pat_i(\pi)$?

\noindent\textbf{Answer:} Each equivalence class in $eqs(p)$ of size
$k$ replaces $k$ distinct variables by a single representative in $p$.
\smallskip

\begin{example}\label{ex:strength}
  Consider the graph pattern $p=[0,0,A,A,1,0,B,C,D,E]$ derived from
  $\pi_v=[3,1,4,2,5]$ and $i=5$ detailed in
  Table~\ref{table:strength}. We have $str(p)=5$ and
  $\sigma=[2,6,4,3]$ the length 4 prefix of $\pi_e$.
  Here, the equations derived from $\sigma$
  introduce the equivalence classes:
  $\set{\{x_1,x_2,x_6\}, \{x_3,x_4\}}$. The equivalence classes
  indicate that 3 variables are eliminated from $p$ (2 because of the
  first equivalence class and 1 because of the second) and so, the
  strength of $p$ is $10 - 2 - 3 =5$.
\end{example}

\medskip\noindent\textbf{Incrementality:~~}
An interesting property of lex-leader canonical graphs, when
lexicographic order is defined in terms of the column-wise
concatenation of edge variables, is the following: 
An order-$n$ canonical graph contains a canonical subgraph on the
first $k$ vertices for every
$1\leq k\leq n$~\cite{Kvasnicka}. As shown
in~\cite{ItzhakovC20}, this property can be exploited
when generating symmetry breaking constraints, as the symmetry break
for order-$n$ graphs can be extended to one for order-$n+1$ graphs.
The following demonstrates how this applies to graph-patterns.

\begin{example}\label{example:incremental-patterns}
  
Observe in Table~\ref{table:strength}, the 9 graph patterns $\pat_i(\pi)$
where $\pi(5)=5$ and $i\leq 6$. These are extensions of the following
9 graph patterns which are a complete symmetry break for order-4
graphs.
\medskip

\begin{tabular}{|lll|lll|}
  \hline
$[3,2,1,4]-1$ & $[1,A,0,B,C,D]$  &\qquad &\qquad  &$[4,3,1,2]-3$ & $[A,0,1,0,B,A]$ \\
$[3,1,2,4]-1$ & $[1,0,A,B,C,D]$  &  &&$[2,4,1,3]-3$ & $[A,A,1,0,A,B]$ \\
$[1,2,4,3]-2$ & $[A,1,B,0,C,D]$  &  &&$[3,1,2,4]-4$ & $[A,A,A,1,B,0]$ \\
$[2,3,1,4]-2$ & $[0,1,0,A,B,C]$  &  &&$[3,1,4,2]-5$ & $[0,0,A,A,1,0]$ \\
$[2,1,3,4]-4$ & $[A,B,B,1,0,C]$  &&&&\\
\hline                                     
\end{tabular}
\end{example}

\medskip\noindent\textbf{CEGAR:~~}
In~\cite{Itzhakov2016} and in~\cite{CodishImps} the authors compute
complete symmetry breaks for graphs which are based on permutations
and on a refinement of permutations which they term ``implications,''
respectively. A similar approach is applied in~\cite{danco2025} to
break the symmetries for finite models.
In~\cite{CodishJanotaCP25}, the authors adapt this approach to compute
symmetry breaks consisting of graph patterns. Common to all of these
works is an algorithm based on \emph{counter-example guided
  abstraction refinement (CEGAR)}~\cite{cegar2000}.
In a nutshell, in the context of graph patterns, the CEGAR-based
algorithm performs as follows: Let $\Psi$ denote a set of graph
patterns, which is initially empty. The algorithm repeatedly seeks a
counter-example to the statement: ``\emph{$\Psi$ is a complete
  symmetry break}''. A counter-example takes the form: a graph $G$ and
a graph pattern $\pat_i(\pi)$ such that $G$ is covered by
$\pat_i(\pi)$ but is not covered by any of the graph patterns in
$\Psi$. If such a counter-example is found then
$\Psi=\Psi\cup\{\pat_i(\pi)\}$. If no such counter-example is found,
then $\Psi$ is a complete symmetry break.  The search for 
counter-examples is implemented using a SAT encoding and incremental
SAT solving.

To encode that $G$ is covered by a pattern $\pat_i(\pi)$, denoting
$edges(G)=\tuple{x_1,\ldots,x_m}$  and
$edges(\pi(G))=\tuple{y_1,\ldots,y_m}$, 
we introduce clauses to capture that
\begin{equation}
  \label{eq:cover}
  (x_1=y_1) \land \cdots \land
  (x_{i-1}=y_{i-1}) \land (x_i=1) \land (y_i=0).
\end{equation}
Each equation $(x_i=y_i)$ is encoded in the CNF so that the atom
$(x_i=y_i)$  can be viewed as a
propositional variable.
To encode that $G$ is not covered by the pattern $\pat_i(\pi)$, we
take the negation of Equation~\eqref{eq:cover} which, given the
encodings of the equations $(x_i=y_i)$, is a single CNF clause.
In this paper we make extensive use of a CEGAR-based algorithm.

\section{Approximating the Strength of Graph Patterns}

The aim in this paper is devise a SAT based approach to construct
partial symmetry breaks by repeatedly selecting graph patterns,
strongest first. However, a SAT encoding to specify the strength of an
unknown graph pattern $p$ is complex due to the need to capture the
equivalence classes of the equations $eqs(p)$. This requires an
encoding of the transive closure of the unknown set of equations which
is costly.
As an alternative, we encode an approximation of $str(p)$. Underlying
the approach is an analysis of a prefix of the edge permutation which
governs the specification of these equations.
To this end, we analyze the positions $j$ in the domain of $\pi_e$
which occur before $i$ and reason about how they impact the strength
of the pattern.  We first demonstrate with an example.
\begin{example}\label{example:approx_strength}
  Consider again, the graph pattern $p=[0,0,A,A,1,0,B,C,D,E]$ of
  strength 5 detailed in Table~\ref{table:strength} which derives from
  the vertex permutation $\pi_v=[3,1,4,2,5]$ with index $i=5$ (see
  also Example~\ref{ex:strength}).
  The length 4 prefix, $[2,6,4,3]$ of the corresponding edge
  permutation gives rise to 4 equations:
  equations $x_3{=}x_4$ and $x_4{=}x_3$ involve a pair of variables
  prior to $i=5$ and corresponds to a swap in $\pi_e$; equation
  $x_1{=}x_2$ involves a pair of variables prior to $i$ which is not a
  swap, and equation $x_2{=}x_6$ involves a single variable prior to $i$.
\end{example}

Let $p=\pat_i(\pi_v)$ be a grah pattern, let
$\sigma$ be the length $i-1$ prefix of a permutation $\pi_e$ which
corresponds to $\pi_v$.
We distinguish between four types of values in $\sigma$.
We say that values $j$ and $k$ in $\sigma$  are
\emph{pre-$i$-swap} if such that $j<k<i$, $\pi_e(j)=k$ and
$\pi_e(k)=j$.
The value $j<i$ in $\sigma$ is \emph{pre-$i$-lower} if $\pi_e(j) < i$
and $\pi_e(\pi_e(j)) \neq j$.
The value $j<i$ in $\sigma$ is \emph{pre-$i$-upper} if $\pi_e(j) > i$.
All other values $j<i$ in $\sigma$ are \emph{pre-$i$-fixed-point},
$\pi(j)=j$.

Each pair of pre-$i$-swap values $j<k$ in $\pi_e$ introduces exactly
one equivalence (between the variables $x_j$ and $x_k$) and weakens
the value of $str(p)$ by one.
Each pre-$i$-upper value $j$ in $\pi_e$
introduces exactly one equivalence (between $x_j$ and a variable with
index greater
than $i$) and weakens the value of $str(p)$ by one.
Each pre-$i$-fixed-point value in $\pi_e$ introduces
zero equivalences and does not influence the  value of $str(p)$.
A pre-$i$-lower value $j<i$ in $\pi_e$ introduces at most one
equivalence, depending on the size of the path or cycle in which $j$
resides.

 \begin{table}
   \caption{Graph patterns for order-5 graphs with corresponding strengths and types of
     positions (note for swaps we count pairs).} 
   \label{table:strength2}
   \centering
 \begin{tabular}{cccccc}
   \hline
   $\pi-i$  &  $p=\pat_i(\pi)$ & $str(p)$ & pre-swaps & pre-lower & pre-upper \\
   \hline
   $[3,2,1,4,5]-1$ & $[1,A,0,B,C,D,E,F,G,H]$ & $8$ & $0$ & $0$ & $0$ \\
   $[3,1,2,4,5]-1$ & $[1,0,A,B,C,D,E,F,G,H]$ & $8$ & $0$ & $0$ & $0$ \\
   $[1,2,4,3,5]-2$ & $[A,1,B,0,C,D,E,F,G,H]$ & $8$ & $0$ & $0$ & $0$ \\
   $[1,2,3,5,4]-4$ & $[A,B,C,1,D,E,0,F,G,H]$ & $8$ & $0$ & $0$ & $0$ \\
   \hline
   $[3,5,4,2,1]-2$ & $[A,1,B,C,D,0,E,F,A,G]$ & $7$ & $0$ & $0$ & $1$ \\
   $[2,3,1,4,5]-2$ & $[0,1,0,A,B,C,D,E,F,G]$ & $7$ & $0$ & $0$ & $1$ \\
   $[2,1,3,4,5]-4$ & $[A,B,B,1,0,C,D,E,F,G]$ & $7$ & $1$ & $0$ & $0$ \\
   $[1,2,3,5,4]-5$ & $[A,B,C,D,1,E,D,0,F,G]$ & $7$ & $0$ & $0$ & $1$ \\
   \hline
   $[4,3,1,2,5]-3$ & $[A,0,1,0,B,A,C,D,E,F]$ & $6$ & $0$ & $0$ & $2$ \\
   $[2,4,1,3,5]-3$ & $[A,A,1,0,A,B,C,D,E,F]$ & $6$ & $0$ & $1$ & $1$ \\
   $[3,1,2,4,5]-4$ & $[A,A,A,1,B,0,C,D,E,F]$ & $6$ & $0$ & $3$ & $0$ \\
   $[2,1,4,3,5]-7$ & $[A,B,C,C,B,D,1,0,E,F]$ & $6$ & $2$ & $0$ & $0$ \\
   $[1,3,2,4,5]-8$ & $[A,A,B,C,D,D,E,1,0,F]$ & $6$ & $2$ & $0$ & $0$ \\
   $[1,2,4,3,5]-9$ & $[A,B,C,B,C,D,E,F,1,0]$ & $6$ & $2$ & $0$ & $0$ \\
   \hline
   $[3,1,4,2,5]-5$ & $[0,0,A,A,1,0,B,C,D,E]$ & $5$ & $1$ & $1$ & $1$ \\
   $[2,1,4,3,5]-9$ & $[A,B,C,C,B,D,E,E,1,0]$ & $5$ & $3$ & $0$ & $0$ \\
   \hline
   $[2,3,5,1,4]-5$ & $[A,0,A,A,1,B,C,0,A,D]$ & $4$ & $0$ & $2$ & $2$ \\
   $[3,5,1,2,4]-6$ & $[0,A,B,B,C,1,B,C,0,D]$ & $4$ & $0$ & $1$ & $3$ \\
   \hline
 \end{tabular}
 \end{table}

 Table~\ref{table:strength2} enhances the graph patterns detailed in
 Table~\ref{table:strength} with the number of positions of three
 types: pre-swap, pre-lower, and pre-upper. All other positions are
 pre-fixed-point positions.
Observe in the table that for all of the rows except that for the
pattern $[3,1,2,4,5]-4$ we have
\[str(p) = \binom{n}{2}-2 - \#\mbox{pre-swaps} - \#\mbox{pre-lower}-
  \#\mbox{pre-upper}\]
We take the approach that, minimizing the values of
$(\#\mbox{pre-swaps} + \#\mbox{pre-lower} + \#\mbox{pre-upper})$ is an
approximation of maximizing the value of $str(p)$.

 \begin{definition}[$\equivs_n(s_1,s_2,s_3)$] \label{defs:equivs}
  Let $\equivs_n(s_1,s_2,s_3)$ denote the set of order-$n$ graph patterns
  $p$, such that the number of pre-swap (pairs), and the numbers of
  pre-lower and pre-upper (values) in $p$ are at most $s_1,s_2$,
  and $s_3$ respectively. When clear from the context we omit the
  subscript $n$. If pattern $p$ belongs to $\equivs_n(s_1,s_2,s_3)$, we
  say that $p$ is $\equivs_n(s_1,s_2,s_3)$.
 \end{definition}

\medskip\noindent\textbf{Encoding  $\equivs_n(s_1,s_2,s_3)$:~~}
Let $n$, $i$, and $s_1,s_2,s_3$ be given integer values such that
$1\leq i,s_1,s_2,s_3 \leq m$ where $m={n\choose 2}$.
The SAT encoding that specifies that $\pi_v$ is an order-$n$
permutation such that $\pi_v-i$ corresponds to a graph pattern in
$\equivs_n(s_1,s_2,s_3)$ is detailed in
Equations~(\ref{eq:encode1}---\ref{eq:encode3}).
In our implementation, the constraints from these equations are
compiled to CNF using the finite-domain constraint compiler
\bee~\cite{jair2013}.

\medskip\noindent \underline{Step one (\emph{channelling})}.~~ First
we express in Equation~\eqref{eq:encode1} that $\pi_v$ and $\pi_e$ are
corresponding vertex and edge permutations.
Both permutations are represented as lists of integer variables which
are constrained to be all different.
\begin{equation}
  \label{eq:encode1}
   \begin{array}{ll}
    \pi_v = [p_1,\ldots,p_n]            & \pi_e = [q_1,\ldots,q_m]\\
    \bigwedge\limits_{i=1}^n integer(p_i,1,n)  & \bigwedge\limits_{i=1}^m integer(q_i,1,m)\\
    \mathtt{allDifferent} (p_1,\ldots,p_n)
                    \qquad\qquad\qquad & \mathtt{allDifferent} (q_1,\ldots,q_m)
  \end{array}
\end{equation}
The integer variables in $\pi_v$ specify a mapping
$[1,\ldots,n]\mapsto [p_1,\ldots,p_n]$ (the $p_i$ are variables). We
constrain $\pi_e$ to express a corresponding mapping
$[1\ldots,m]\mapsto [q_1,\ldots,q_m]$. To this end, denote
$id_e = [(1,2),(1,3),(2,3),\ldots,(n-1,n)]$. This specifies the edge
positions in the column-wise order. So, $q_i=j$ if $\pi_v$ maps the
$i^{th}$ element of $id_e$ to the $j^{th}$ element of $id_e$ as
specified by the constraints in Equation~\eqref{eq:encode2}.
\begin{equation}
  \label{eq:encode2}
  \begin{array}{lll} 
    \bigwedge\limits_{i=1}^m \bigwedge\limits_{j=1}^m~~
    q_i = j &\leftrightarrow& id_e[i]=(u_i,v_i) ~\land~
                              id_e[j]=(u_j,v_j) ~\land~ \\
            && (p_{u_i} = u_j ~\land~ p_{v_i}=v_j) ~\lor~
               (p_{u_i} = v_j ~\land~ p_{v_i}=u_j)
  \end{array}
\end{equation}

\medskip\noindent \underline{Step two
  (\emph{$\equivs(s_1,s_2,s_3)$})}.~~ Now we encode the parameters
$s_1,s_2,s_3$ corresponding to the numbers of pre-swaps, pre-lower and
pre-upper positions in the graph pattern corresponding to
$\pi_v-i$. Let $\pi_e=[q_1,\ldots,q_m]$.

\begin{equation}
  \label{eq:encode3}
  \begin{array}{ll}
  \bigwedge\limits_{a<b<i}x_{a,b}\leftrightarrow~~q_a=b~\land~q_b=a 
  \qquad\qquad\qquad & \sum\limits_{a<b<i} x_{a,b}\leq  s_1 \\[4ex]
  \bigwedge\limits_{a\leq b\leq i}y_{a,b}\leftrightarrow~~q_a=b~~\mathtt{xor}~~q_b=a  
                     & \sum\limits_{a<b<i} y_{a,b}\leq  s_2 \\[4ex]
 \bigwedge\limits_{a<i}z_a\leftrightarrow~~ q_a>i
                     & \sum\limits_{a<i} z_a \leq  s_3   
  \end{array}
\end{equation}

\section{Strength-driven, Redundancy-aware Pattern Generation}

We define a series of partial symmetry breaks
$\Psi^n_1,\ldots,\Psi^n_k$ for order-$n$ graphs defined in terms of
sets of graph patterns $\ell^n_1,\ldots,\ell^n_k$ where $\ell^n_1$ is
the set of all order-$n$ transpositions and for $j>1$,
$\ell^n_j= \equivs_n(s^j_{1},s^j_{2},s^j_{3})$.
The sequence of partial symmetry breaks can be viewed as a sequence of
layers: each element is computed ``on top'' of the previous element.

To compute the symmetry breaks, we start from an empty set of graph
patterns, $\Psi^n_0=\emptyset$ and apply in iteration, where $j$ takes
values from $1$ to $k$:
\[\Psi^n_j = \mathtt{generate(\ell^n_j)}\]
The procedure $\mathtt{generate(\ell^n_j)}$ is CEGAR-based.  It
initializes $\Psi^n_j$ with the graph patterns from the previous layer
$\Psi^n_{j-1}$ and then, so long as there exists a counter-example from
$\ell^n_j$ to the fact that $\Psi^n_j$ is not a complete symmetry
break, it adds the corresponding graph pattern to $\Psi^n_j$.

We note that when the computation of $\Psi^n_j$ terminates, $\Psi^n_j$
is most likely not a complete symmetry break. However, no
counter-example from $\ell^n_j$ exists. We also note that the set of
graph patterns, $\ell^n_j$, is not represented explicitly. It is
represented as a set of constraints (a propositional formula).

The idea is to design the sequence $\ell^n_1,\ldots,\ell^n_k$ so that
stronger graph patterns are selected first and also to make layered
based computation feasible. Note that we could simply define the
sequence $\tuple{\ell^n_k}$ or the sequence
$\tuple{\ell^n_1,\ell^n_k}$ but this does not work computationally.
The specific sequence considered in our experimentation is the
following:
\[
  \begin{array}{l}
       trans,\equivs(3,2,2),\equivs(4,3,3),\equivs(5,3,2),\equivs(6,3,3),\equivs(7,4,3),\\
       \equivs(8,4,4),\equivs(9,5,4),\equivs(10,5,4),\equivs(11,6,5),\equivs(12,6,5),\\
    \equivs(13,6,6),\equivs(14,7,6),\equivs(15,7,7),\equivs(16,8,7),\equivs(17,8,8),\\
    \equivs(18,9,9),\equivs(19,10,9),\equivs(20,10,10).
  \end{array}
\]
However, for values of $n>14$, we only have  calculated the symmetry breaks
up to $\equivs(17,8,8)$.
The general approach described above is the basis of our method, but in
order to ensure efficient execution we need to make use of additional
heuristics and considerations.

\medskip\noindent\textbf{Initialization (and incrementality):~~}
In fact, because we have chosen to order the edge variables by column,
we can do better at the initialization phase in the definition of
$\mathtt{generate(\ell^n_j)}$. For $n>4$, we initialize
$\Psi^n_j=\Psi^n_{j-1} \cup extend(\Psi^{n-1}_{j})$ where each graph
pattern of order-$n-1$ is extended to one of order-$n$ (by
concatenating $n-1$ fresh variables)~\cite{ItzhakovC20}. See also
Example~\ref{example:incremental-patterns}.

\medskip\noindent\textbf{An inner layer of iteration (and a heuristic):~~}
In the implementation, when iterating over layers, when we compute
$\mathtt{generate(\ell^n_j)}$, we iterate over all values
$1\leq i\leq \binom{n}{2}$ where $i$ is the index of the
corresponding graph patterns $\pat_i(\pi)$ considered from
$\ell^n_j$. This simplifies the encoding of ``is there a counter-example
from $\ell^n_j$''. We encode, for each $i$: ``is there a
counter-example of the form $\pat_i(\pi)$ from $\ell^n_j$'' for the
given $i$.
In practice, especially for larger values of $n$, the majority of indices 
$1\leq i\leq \binom{n}{2}$ do not lead to counter-examples (do not
add new graph patterns). In practice, this leads to a large number of
expensive SAT calls which are ``unsat''.
We observe for $n\leq 10$ that there are
only a few select values of $i$ for which there exist counter-examples
of the form $\pat_i(\pi)$ from $\ell^n_j$. Moreover, the set of such
indices remains, more or less, constant in $\ell^n_j$ as $n$ increases.
Hence, we introduce a heuristic: First we determine which values of
$i$ contribute graph patterns in the computations for smaller values
$n\leq 10$. Then, for larger values of $n>10$ we restrict the search
to these specific values of $i$.
It means that we might not be adding all graph patterns that could
otherwise be detected at layer $\ell^n_j$.

\medskip\noindent\textbf{Keeping the symmetry breaks ``small'' (a heuristic):~~}
For the specific layers in our design, the number of graph patterns
added at each iteration $i$ for layer $\ell^n_j$ is usually small (less than hundreds).
In some cases it is  huge (thousands, tens of thousands).
As a heuristic, we restrict the number of graph patterns added at each
iteration $i$ for layer $\ell^n_j$ to 1000. However, we do not take the
first 1000 counter-examples. We take the last 1000
counter-examples. These have proven non-redundant given all of the
previous ones.

\medskip\noindent\textbf{The reduce procedure:~~}
An important detail regarding CEGAR-based algorithms is that $\Psi$
may contain redundant elements. For example, if a graph pattern added at
some point becomes redundant in view of those added later. A
second phase of the algorithm iterates on the elements of $\Psi$ to
remove redundant permutations (similar to the iterative algorithm for
a minimally unsatisfiable set or monotone predicates in
general~\cite{humus,monotone}). It is important to note that the time
to perform the second phase is costly.
In our implementation we apply the reduce algorithm between layers and
also periodically inside layers (every $n^2\times 40$ graph patterns
derived).

\medskip\noindent\textbf{On the selection of the specific layers:~~}
The specific values of $s_1$, $s_2$ and $s_3$ chosen in the layers are
the result of trial and error. We did not start from $\equivs(1,1,1)$
as this resulted in very few graph patterns. Clearly a more principled
investigation of the sequence is warranted.

\medskip\noindent\textbf{On the strategy:~~}
It is important to note that selecting graph patterns ``strongest
first'' is not guaranteed to make the best choice.
Given a set of graph patterns already selected, a strong graph pattern
may cover only a small number of additional graphs.

\medskip\noindent\textbf{Summary:~~}
We call our approach ``strength-driven'', as we select patterns with
decreasing strength. We call it ``redundancy-aware'' because the
CEGAR-based algorithm ensures that every graph pattern selected is not
redundant (at the time it is generated).

\medskip Table~\ref{table:ten} details the sizes and redundancy ratios
for several of the symmetry breaks generated for order-10 graphs. On
the left, transpositions which are currently the most widely applied
partial symmetry break (with redundancy ratio 15.34).  On the right, a
complete symmetry break (with redundancy ratio 1.00).  Notice that
using $\equivs(10,5,4)$ we obtain a symmetry break with less than 10\% of
the graph patterns required for a complete symmetry break with a
redundancy ratio of 1.34. This means that when using this symmetry
break we will obtain on average 1.34 graphs for every canonical
solution.

\begin{table}
   \caption{Partial and complete symmetry breaks for order-10 graphs.} 
   \label{table:ten}
   \centering
{%
\begin{tabular}{|c|c|c|c|c|c|c|c|c|c|c|}
\hline
  & trans & equivs(10,5,4)
    & equivs(14,7,6)   & equivs(15,7,7)   &  equivs(17,8,8) & complete \\
\hline
  size  & 128   & 2273 & 3300 & 5802 & 7471 & 25263 \\
\hline
 $\rho$ & 15.34 & 1.34& 1.21 & 1.14 & 1.07 & 1.00 \\
\hline
\end{tabular}
}
 \end{table}

\section{Our Partial Symmetry Breaking Constraints}

This section details properties of the partial symmetry breaks
constructed: their precision (redundancy ratios), the times required to
compute them, and their sizes.

\begin{figure}[htbp]
     \centering
     \begin{subfigure}[b]{0.49\textwidth}
         \centering
         \includegraphics[width=\textwidth]{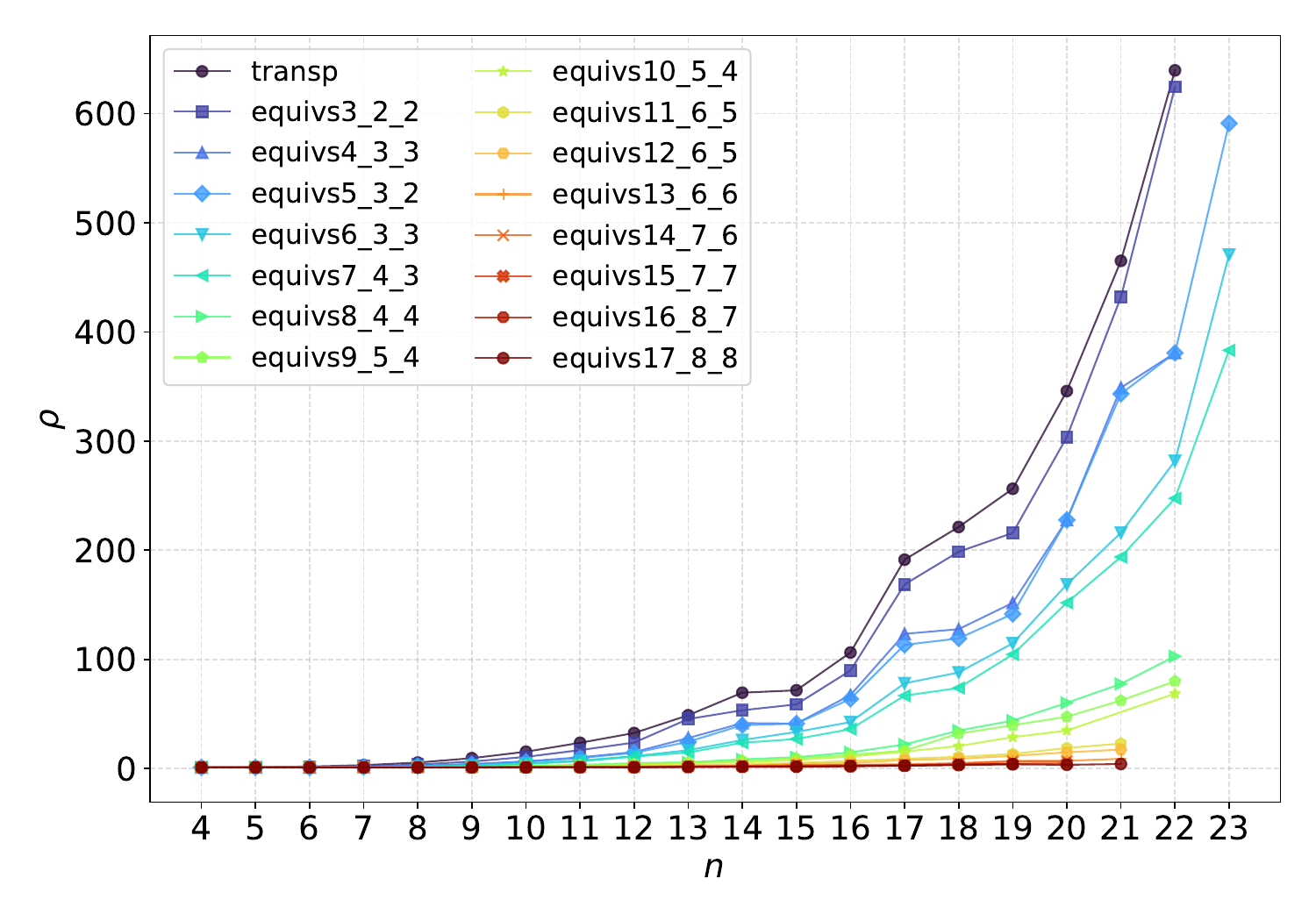}
         \caption{$\rho$ vs $n$ (Grouped by Type)}
         \label{fig:rbytype}
     \end{subfigure}
     \hfill
     \begin{subfigure}[b]{0.49\textwidth}
         \centering
         \includegraphics[width=\textwidth]{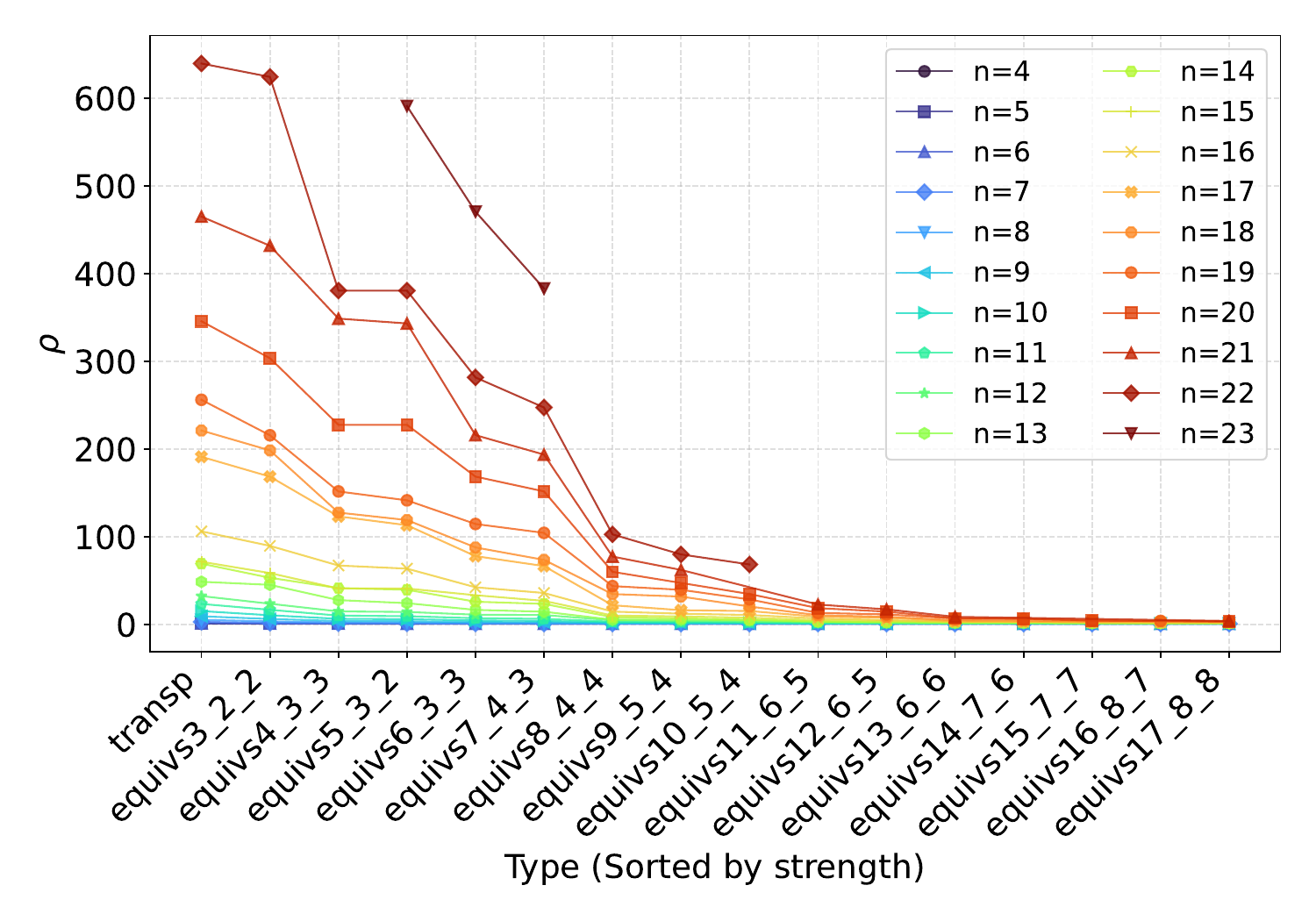}
         \caption{$\rho$ vs Type (Grouped by $n$)}
         \label{fig:rbyn}
     \end{subfigure}
     \caption{Values of $\rho$ for different symmetry breaks and values of $n$.}
     \label{fig:ratios}
\end{figure}

Figure~\ref{fig:ratios} shows redundancy ratios~$\rho$ for the different
symmetry breaks and different values of~$n$ (number of vertices). The
redundancy ratio is computed as the ratio between the number of graphs
which are solutions of each corresponding symmetry break and the
number of non-isomorphic graphs. The numerator is computed by calling
a model counter to compute the number of graphs which are solutions of
each corresponding symmetry break, and the value of the denominator is
found as~\cite{oeis:A000088}. When possible with a 4 hour timeout
the ratio is computed using an exact model counter
\texttt{ganak-2.5.2}~\cite{ganak}, otherwise it
is computed using an approximate model counter
\texttt{approx-4.2}~\cite{approxmodelcounter}.  Full details on the
values of the ratio are given in
Appendix~\ref{appendix:symBreaks}.

Figure~\ref{fig:rbytype} shows how the values evolve for each
symmetry break for different values of $n$. Figure~\ref{fig:rbyn} shows how the
values evolve as the symmetry break gets stronger.
In Figure~\ref{fig:rbytype}, the upper curve corresponds to the use of
transpositions, as in~\cite{CodishMPS13,Codish2019}. The lower curve
corresponds to our most precise symmetry break. This figure makes it clear
that the approach taken to derive symmetry breaks,
``strongest-first'', results in symmetry breaks 
which evolve to give good precision. 
Missing points in the plots (for instance in the upper plot of
Figure~\ref{fig:rbytype}) indicate that the corresponding SAT instances were
too hard for the approximate model counter, so we cannot compute $\approx
\rho$. 
Similarly, although both
plots in Figure~\ref{fig:ratios} detail values up till $n=23$, symmetry
breaks have been computed up to 25.  In all cases we can compute the
symmetry breaks and apply them.

\begin{figure}[htbp]
     \centering
     \begin{subfigure}[b]{0.49\textwidth}
         \centering
         \includegraphics[width=\textwidth]{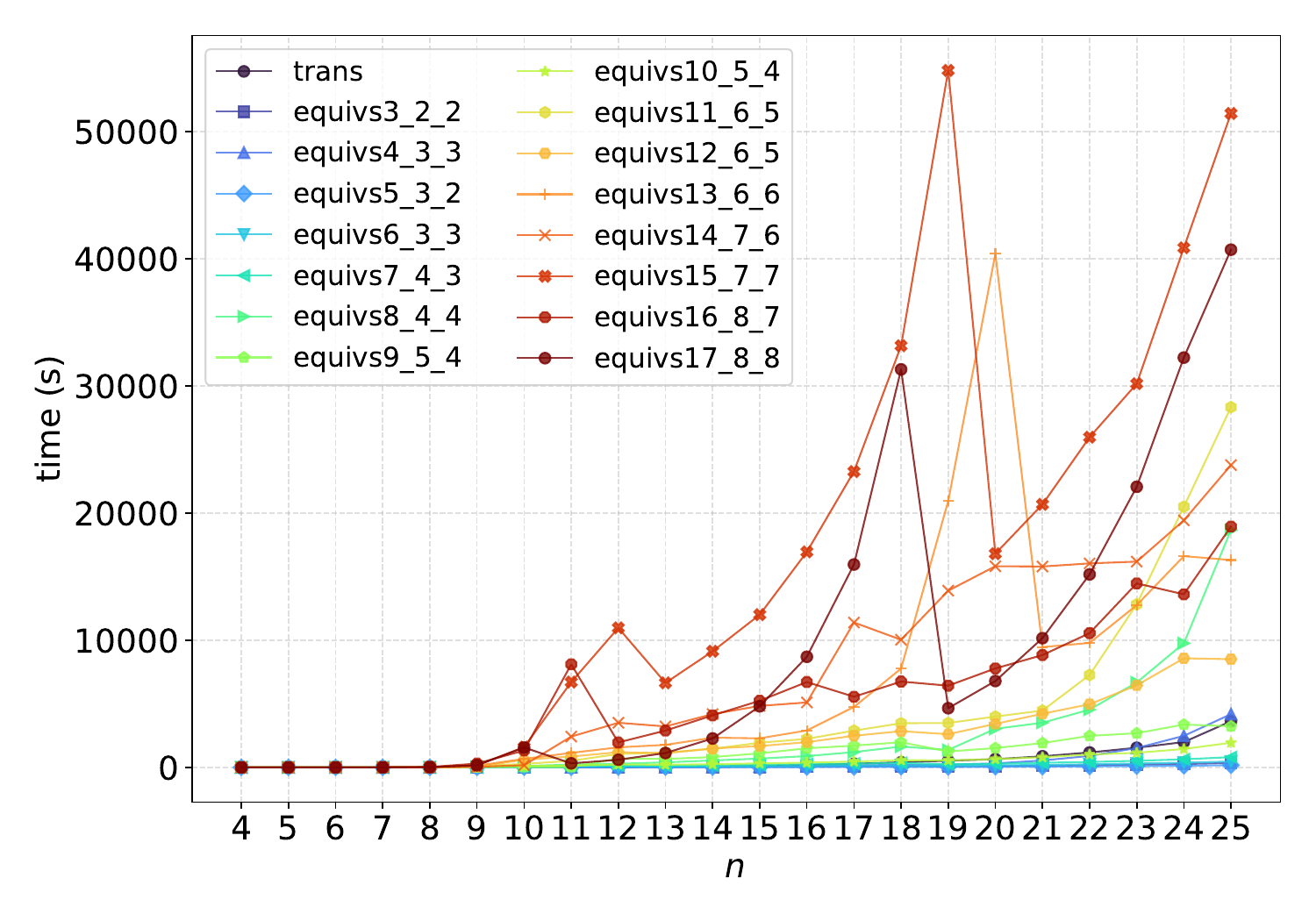}
         \caption{Time vs n (Grouped by Type)}\label{fig:tbytype}
     \end{subfigure}
     \hfill
     \begin{subfigure}[b]{0.49\textwidth}
         \centering
         \includegraphics[width=\textwidth]{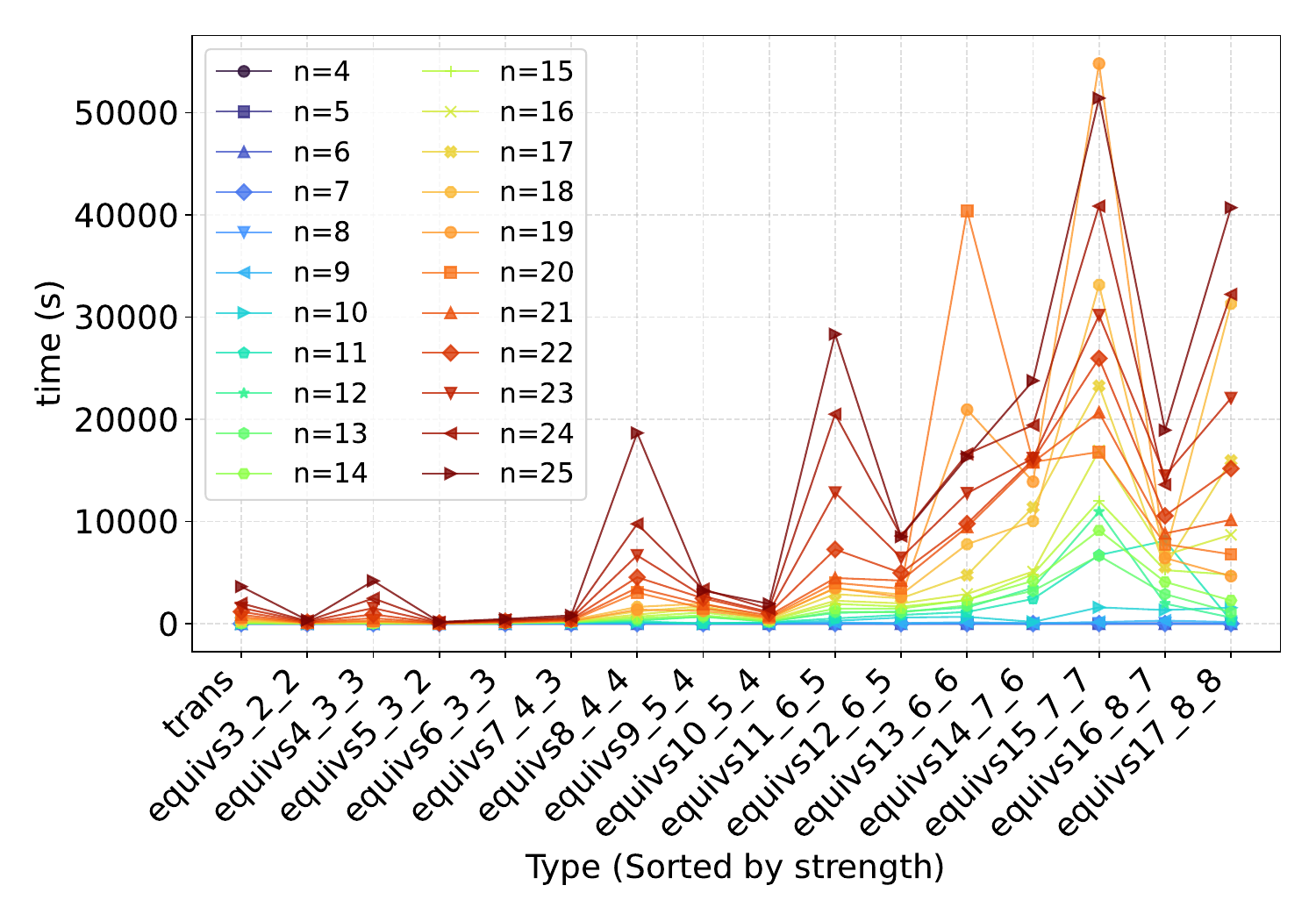}
         \caption{Time vs Type (Grouped by n)}\label{fig:tbyn}
     \end{subfigure}
     \caption{Time to calculate the sym.\ break for different symmetry breaks and values of n.}
     \label{fig:times}
\end{figure}

\begin{figure}[htbp]
     \centering
     \begin{subfigure}[b]{0.49\textwidth}
         \centering
         \includegraphics[width=\textwidth]{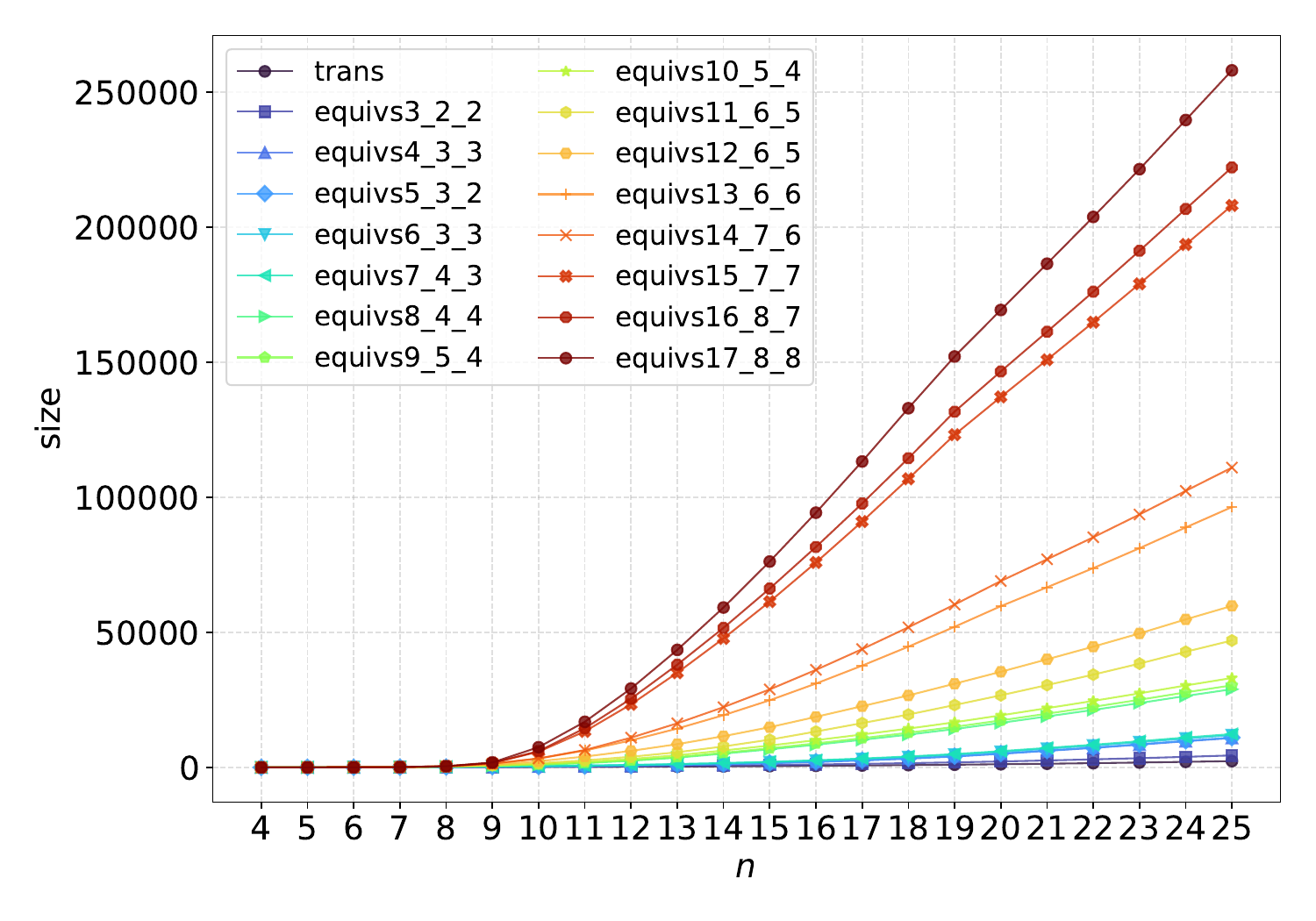}
         \caption{size vs n (Grouped by Type)}\label{fig:szbytype}
     \end{subfigure}
     \hfill
     \begin{subfigure}[b]{0.49\textwidth}
         \centering
         \includegraphics[width=\textwidth]{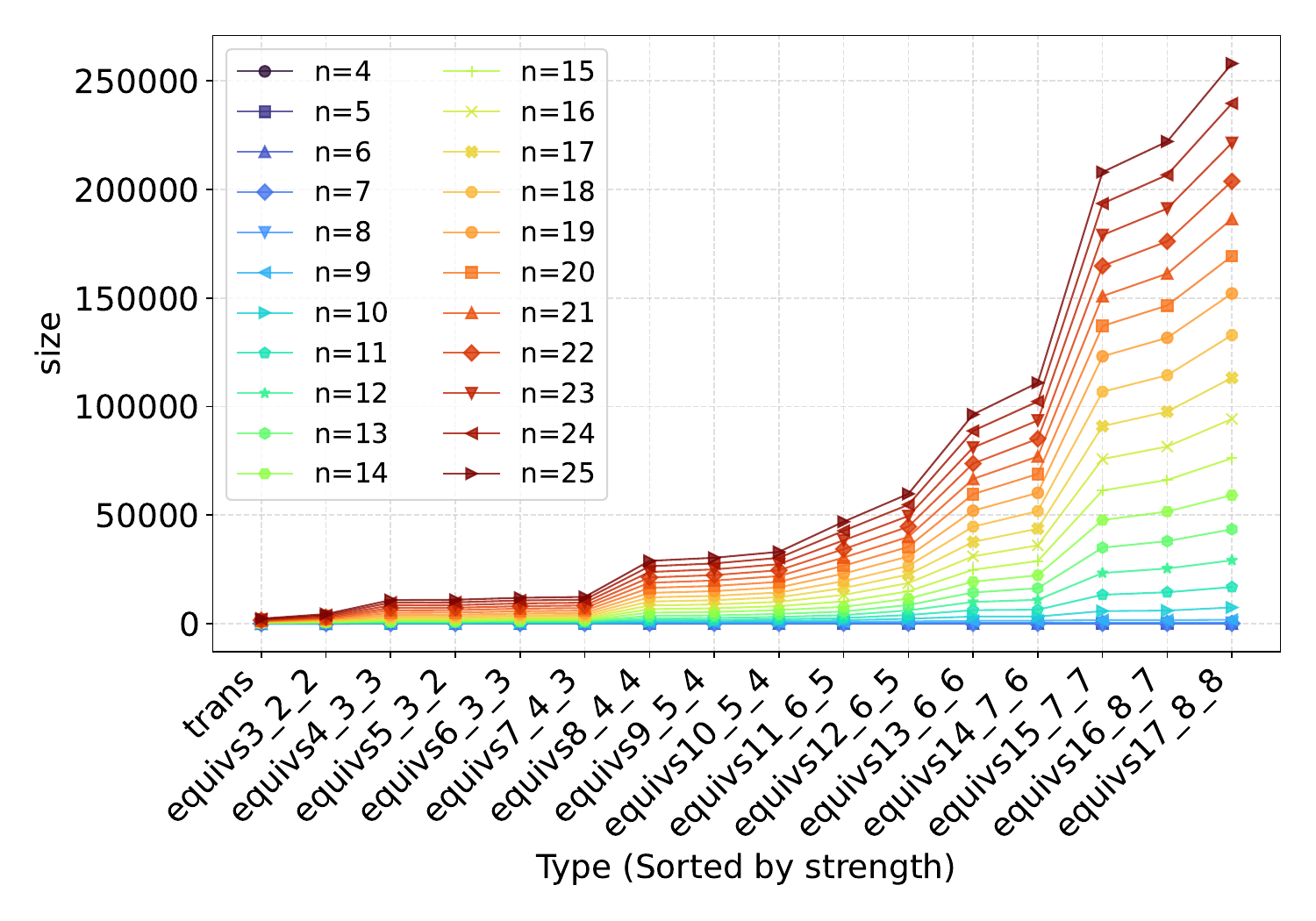}
         \caption{size vs Type (Grouped by n)}\label{fig:szbyn}
     \end{subfigure}
     \caption{The size of the sym.\ break for different symmetry breaks and values of n.}
     \label{fig:sizes}
\end{figure}

Figure~\ref{fig:times} shows the \emph{incremental} times needed to 
produce the symmetry breaks for different values of $n$ (number of vertices) for
each layer---that is this is the time to complete the layer $\Psi^n_j$
starting from $\Psi^n_{j-1} \cup extend(\Psi^{n-1}_{j})$.

Figure~\ref{fig:tbytype} shows how the time evolves for each symmetry break for
different values of $n$. 
While usually the time grows with both $n$ and the layer number $j$, it is
not monotonic in either. This is potentially a result of the differences between
subsequent layers not being uniform. 
Figure~\ref{fig:tbyn}
shows how the time evolves as the symmetry break gets stronger. 
Clearly in general while time required grows with layers it is not uniform,
again as a result of the fact that the difference between subsequent layers
are not uniform.
Many of the peaks in Figure~\ref{fig:tbyn} correlate to
the number of indices considered by the relevant indices heuristic.
Analogously, Figure~\ref{fig:sizes} shows the sizes of the  symmetry breaks.
Full details on the sizes and times to compute the proposed symmetry
breaks is given as Appendix~\ref{appendix:symBreaks}.

\section{Evaluation}\label{sec:eval}

We consider two benchmark problems and illustrate the impact of our
partial symmetry breaks when solving them. In both cases we compare to
the partial symmetry break defined in terms of transpositions and to
the use of dynamic symmetry breaking using~\cite{sms}.
For both benchmarks the encoding of the problem instances to SAT is
performed using the SMS tool.\footnote{Obtained from
  \url{https://github.com/markirch/sat-modulo-symmetries} under
  revision \texttt{cf890fd}, using the tool's options
  \texttt{\textminus\textminus no-solve \textminus\textminus cnf FILE}
  to generate a CNF file.} 
Then we couple the resulting CNF with
other symmetry breaks. %
All the experiments were performed on a machine with two AMD EPYC 7513 32-Core
processors and with 514~GiB~RAM with 100 problem instances run in parallel with
4h timeout.

\medskip\noindent\textbf{Diameter-2-Critical Graphs:~~} 
The diameter of a graph $G$ is the largest distance among all pairs of
vertices in $G$, where the distance of two vertices is the length of a
shortest path between them. A disconnected graph has diameter
$\infty$. A graph is diameter-$d$-critical if its diameter is $d$, but the
deletion of any edge increases the diameter.
In~\cite{sms}, the authors report that previous work on this problem
provides results for graph of size $n\leq 11$ and that they were able
to extend these results to graphs with $n\leq 13$ vertices. 
We refer the reader to \cite{sms} for details on the origin and
history of the problem.

\begin{table}
  \caption{Diameter 2 Critical comparison}\label{tab:diam}
  \begin{center}
    \begin{tabular}[c]{||l||r|r||r|r|r||r|r|r||}
      \hline
    $n$ &
      \multicolumn{2}{|c||}{SMS} &
      \multicolumn{3}{|c||}{Static-best}&
      \multicolumn{3}{|c||}{Static Trans.}  \\\hline
   & \textbf{models}  & \textbf{time (s)} &
     \textbf{models}  & \textbf{time (s)} & ratio &
     \textbf{models}  & \textbf{time (s)} & ratio \\\hline
      4    & 2                 & 0.00    & 2                 & 0.00     & 1.00 &  2                   &  0.01  &1.00 \\
      5    & 3                 & 0.00    & 3                 & 0.00     & 1.00 &  4                   &  0.00  &1.33 \\
      6    & 5                 & 0.00    & 5                 & 0.01     & 1.00 &  11                  &  0.01  &2.20 \\
      7    & 10                & 0.00    & 10                & 0.01     & 1.00 &  32                  &  0.01  &3.20 \\
      8    & 30                & 0.02    & 30                & 0.02     & 1.00 &  163                 &  0.02  &5.43 \\
      9    & 103               & 0.11    & 103               & 0.07     & 1.00 &  1018                &  0.12  &9.88 \\
      10   & \numprint{520}    & 0.52    & 533               & 0.42     & 1.03 &  9727                &  1.62  &18.71  \\
      11   & \numprint{3748}   & 2.10    & \numprint{4051}   & 4.18     & 1.08 &  \numprint{133316}   &  32.03 &35.57  \\
      12   & \numprint{40868}  & 23.34   & \numprint{48115}  & 41.47    & 1.18 &  t.o.                &  t.o.  & t.o.\\
      13   & \numprint{688130} & 774.89  & \numprint{934992} & 1088.75  & 1.36 &  t.o.                &  t.o.  & t.o. \\\hline
    \end{tabular}
  \end{center}
\end{table}

Table~\ref{tab:diam} shows results for the Diameter-2-critical problem.  We
compare static symmetry breaks with our best symmetry break and using
a symmetry break based on transpositions and the computation using the
SMS~\cite{sms} tool. For the static symmetry breaks the models were
enumerated by \texttt{CaDiCaL}~\cite{cadical} using a blocking clause
mechanism. Times are given in seconds.
The two columns titled ratio detail the ratio between the number of
models obtained and the actual number of models (found using SMS).
For this experiment we have used partial symmetry breaks
(static-best) based on $\equivs(20,10,10)$. Details on these symmetry
breaks are given in Appendix~\ref{appendix:symBreaks}.
Comparing the two static symmetry breaks, ``Static-best'' is superior
both in precision and in time. The dynamic approach using SMS is
superior to static symmetry breaking.

\medskip\noindent\textbf{Extremal Graphs with Required Girth:~~}
An extremal graph theory problem: What is the maximal number of edges
in a graph with $v$ vertices and no cycles of length $4$ or less
(i.e.\ graphs with girth at least $5$).
Let $f_4(n)$ denote the maximum number of edges in an order-$n$
graph with girth at least 5.  
We consider the same encoding as applied in~\cite{Codish2019} and
in~\cite{sms}. We consider the largest satisfiable instances which
contain $f_4(n)$ edges and the smallest unsatisfiable instances which
contain $f_4(n)+1$ edges. In both cases the problem is partitioned
depending on values of pairs $(\delta,\Delta)$ which denote the
minimal and maximal degrees of vertices in the solutions (if they
exist). These pairs are determined by known constraints related to
the problem. For a satisfiable instance with $n$ vertices and $m$
edges, we take the minimal time over the corresponding pairs.  For an
unsatisfiable instance with $n$ vertices and $m$ edges, we take the
maximal time over the corresponding pairs. This is the convention set
in~\cite{Codish2019} and in~\cite{sms}.
We refer the reader to those papers for details on the encoding.

\begin{table}
  \caption{Extremal Graphs Comparison}\label{tab:exg}
  \begin{center}
    \begin{tabular}[c]{|l|r|r|r|r|r|r|}
      \hline
       n  & \multicolumn{3}{c|}{SAT} & \multicolumn{3}{c|}{UNSAT} \\ \hline
          & SMS                      & Static-best     & Trans      & SMS    & Static-best      & Trans \\ \hline
       15 & 0.16      & 0.97            & 0.13       & \trivial   & \trivial             & \trivial  \\
       16 & 0.20      & 1.01            & 0.14       & 0.52   & 3.72             & 10.85 \\
       17 & 0.49      & 0.93            & 0.23       & \trivial   & \trivial             & \trivial \\
       18 & 0.53      & 2.17            & 0.55       & \trivial   & \trivial             & \trivial \\
       19 & 0.14      & 0.95            & 0.24       & \trivial   & \trivial             & \trivial \\
       20 & 1.80      & 4.26            & 2.46       & \trivial   & \trivial             & \trivial \\
       21 & 0.59      & 2.93            & 0.85       & 0.85   & 7.66             & 236.90 \\
       22 & 3.44      & 3.71            & 3.33       & 3.00   & 147.93           & 2672.83 \\
       23 & 4.49      & 14.01           & 0.05       & 10.79  & 3718.25          & t.o. \\
       24 & 12.38     & 30.87           & 17.54      & \trivial   & \trivial             & \trivial  \\
       25 & 90.34     & 28.37           & 5.18       & 402.31 & t.o.             & t.o.  \\
      \hline
    \end{tabular}
  \end{center}
\end{table}

Table~\ref{tab:exg} shows the time comparison (in seconds) between dynamic
symmetry break (SMS), transpositions, and our best symmetry break
($equivs(17,8,8)$). We consider values of $n\leq 25$, corresponding to
those for which we have constructed static symmetry breaks.  The cells
marked with \trivial\ indicate that an unsat result follows trivially
from theoretical bounds.
The solver Kissat~\cite{kissat} was used to decide the SAT queries.
For satisfiable instances, all three of the techniques are
similar. For the unsatisfiable instances, the stronger static symmetry
break is superior to the use of static symmetry break based on
transpositions. The dynamic approach using SMS is
superior to static symmetry breaking.

\section{Conclusion} 

Symmetry breaking for graph search problems is challenging: ignoring
symmetries leads to vast amounts of redundant work, while removing all
symmetries is impossible for large $n$. Hence partial symmetry breaks are an
important approach to practical symmetry breaking for graph search. 
In this paper we develop a method for creating ``small'' partial symmetry
breaking constraints which nonetheless have low redundancy ratio, and
importantly are not too difficult to compute.
The method substantially improves on earlier methods for partial symmetry
breaking for graph search problems.

The symmetry breaking constraints constructed in this paper are solver
independent. They can be applied when solving graph search problems
via SAT encodings, constraint solvers, or SMT solvers. In addition,
they are problem independent. They can be applied when solving any
graph search problem. It is important to note that while dynamic
symmetry breaking (e.g. SMS~\cite{sms}) has its advantages (as
indicted also by the benchmarks in Section~\ref{sec:eval}), it
requires adapting an existing solver substantially.

\subsubsection*{Acknowledgements}
The research was supported by the European Union under the project
\textsf{ROBOPROX} (reg.\ no.\ CZ.02.01.01/00/22\_008/0004590) and is part of the
\textsf{RICAIP} project that has received funding from the European Union's
Horizon~2020 research and innovation programme under grant agreement No~857306.

\newpage
\appendix

\section{Computing the Partial Symmetry Breaks}
\label{appendix:symBreaks}

Tables~\ref{tab:appendix:part1}--\ref{tab:appendix:part4} detail the
sizes of the partial symmetry breaks (number of graph patterns) and
the computing times required (seconds). Times are incremental. When
computing the order-$n$ symmetry break at layer $\ell_j$ we first
import the two symmetry breaks: (1) $\Psi^{n-1}_j$: order-$n-1$, layer
$\ell_j$, and (2) $\Psi^{n}_{j-1}$: order-$n$, layer $\ell_{j-1}$.

\medskip
Tables~\ref{tab:big:ratios} and~\ref{tab:big:ratios1} detail the redundancy
ratios of our  partial symmetry breaks (in the columns titled ``ratio'').
When possible with a 4 hour timeout, this is computed using an
exact model counter \texttt{ganak-2.5.2}~\cite{ganak}, and in other cases
this is computed using an approximate model counter
\texttt{approx-4.2}~\cite{approxmodelcounter}. In the latter case, the
value is annotated by the symbol ``$\approx$''. The remaining timeouts
(using the approximate model counter) are denoted ``t.o.''
Timeouts indicate that the corresponding SAT instances were too hard
for the approximate model counter.  In all cases we can compute the
symmetry breaks and apply them. Tables~\ref{tab:big:ratios}
and~\ref{tab:big:ratios1} detail the redundancy ratios for $n\leq
23$. For $n>23$ the model counter times out for all instances.

\begin{sidewaystable}
    \centering
    \caption{Computing partial symmetry breaks (part 1).}
    \label{tab:appendix:part1}
\begin{tabular}{||c||r|r||r|r||r|r||r|r||r|r||r|r||}
  \hline
  n &
    \multicolumn{2}{|c||}{trans} &  
    \multicolumn{2}{|c||}{equivs(3,2,2) } &
    \multicolumn{2}{|c||}{equivs(4,3,3)} & 
    \multicolumn{2}{|c||}{equivs(5,3,2)}     &
    \multicolumn{2}{|c||}{equivs(6,3,3)} & 
    \multicolumn{2}{|c||}{equivs(7,4,3)} \\
    \hline
    & time & size &time & size &time & size &time & size &time & size &time & size   \\
  \hline
4 & 0.02 & 6 & 0.03 & 6 & 0.02 & 6 & 0.02 & 6 & 0.03 & 6 & 0.02 & 6 \\
5 & 0.05 & 13 & 0.20 & 16 & 0.1 & 16 & 0.01 & 16 & 0.04 & 16 & 0.01 & 16 \\
6 & 0.16 & 24 & 0.69 & 32 & 0.54 & 39 & 0.21 & 39 & 0.10 & 39 & 0.03 & 39 \\
7 & 0.47 & 40 & 1.85 & 61 & 2.32 & 81 & 0.45 & 80 & 1.52 & 92 & 1.44 & 98 \\
8 & 1.07 & 62 & 3.35 & 101 & 6.51 & 146 & 1.93 & 144 & 3.58 & 172 & 5.59 & 174 \\
9 & 4.01 & 91 & 4.89 & 156 & 9.27 & 233 & 3.06 & 235 & 15.03 & 283 & 28.46 & 285 \\
10 & 7.82 & 128 & 8.16 & 225 & 12.69 & 341 & 4.49 & 349 & 33.24 & 435 & 83.06 & 437 \\
11 & 12.97 & 174 & 12.66 & 313 & 20.63 & 484 & 10.32 & 501 & 45.56 & 636 & 129.75 & 645 \\
12 & 21.61 & 230 & 18.64 & 419 & 31.44 & 664 & 15.37 & 690 & 61.70 & 877 & 147.95 & 909 \\
13 & 37.25 & 297 & 16.45 & 547 & 26.71 & 896 & 12.05 & 931 & 50.50 & 1174 & 127.34 & 1231 \\
14 & 75.27 & 376 & 23.00 & 696 & 43.19 & 1176 & 15.27 & 1220 & 65.63 & 1519 & 157.75 & 1605 \\
15 & 116.81 & 468 & 34.12 & 871 & 68.86 & 1546 & 19.20 & 1599 & 78.83 & 1953 & 199.33 & 2068 \\
16 & 185.14 & 574 & 46.69 & 1073 & 111.34 & 2009 & 26.02 & 2071 & 103.96 & 2480 & 249.22 & 2624 \\
17 & 285.28 & 695 & 66.64 & 1304 & 146.47 & 2569 & 33.53 & 2640 & 136.17 & 3105 & 293.04 & 3278 \\
18 & 400.41 & 832 & 90.11 & 1566 & 205.33 & 3238 & 45.02 & 3318 & 164.16 & 3838 & 350.22 & 4040 \\
19 & 522.77 & 986 & 70.65 & 1862 & 215.33 & 4037 & 36.88 & 4126 & 125.04 & 4701 & 240.17 & 4932 \\
20 & 645.46 & 1158 & 97.12 & 2191 & 309.61 & 5022 & 45.83 & 5120 & 160.62 & 5750 & 285.49 & 6010 \\
21 & 869.37 & 1349 & 122.11 & 2558 & 541.05 & 6125 & 59.00 & 6232 & 191.46 & 6917 & 363.49 & 7206 \\
22 & 1172.85 & 1560 & 162.98 & 2964 & 902.47 & 7204 & 74.04 & 7328 & 240.57 & 8068 & 423.97 & 8386 \\
23 & 1541.74 & 1792 & 209.77 & 3411 & 1521.21 & 8408 & 96.82 & 8543 & 303.38 & 9338 & 515.48 & 9685 \\
24 & 1990.19 & 2046 & 263.49 & 3900 & 2476.87 & 9681 & 123.13 & 9832 & 373.52 & 10683 & 632.80 & 11059 \\
25 & 3628.05 & 2323 & 343.95 & 4435 & 4185.22 & 10863 & 171.42 & 11023 & 458.08 & 11930 & 806.07 & 12335 \\
\hline
\end{tabular}
\end{sidewaystable}

\begin{sidewaystable}
    \centering
    \caption{Computing partial symmetry breaks (part 2).}
    \label{tab:appendix:part2}
\begin{tabular}{||c|||r|r||r|r||r|r||r|r||r|r||}
  \hline
  n &
    \multicolumn{2}{|c||}{ equivs(8,4,4) } &
    \multicolumn{2}{|c||}{ equivs(9,5,4)} & 
    \multicolumn{2}{|c||}{ equivs(10,5,4)}     &
    \multicolumn{2}{|c||}{ equivs(11,6,5)} & 
    \multicolumn{2}{|c||}{ equivs(12,6,5)} \\
    \hline
    &time & size &time & size &time & size &time & size &time & size   \\
  \hline
4 & 0.02 & 6 & 0.01 & 6 & 0.02 & 6 & 0.02 & 6 & 0.01 & 6 \\
5 & 0.12 & 16 & 0.00 & 16 & 0.01 & 16 & 0.15 & 16 & 0.01 & 16 \\
6 & 0.46 & 39 & 0.01 & 39 & 0.03 & 39 & 0.46 & 39 & 0.02 & 39 \\
7 & 1.94 & 109 & 0.02 & 109 & 0.06 & 109 & 2.70 & 112 & 0.06 & 112 \\
8 & 12.83 & 259 & 3.90 & 288 & 4.57 & 298 & 15.01 & 343 & 0.21 & 343 \\
9 & 45.53 & 535 & 12.29 & 595 & 73.24 & 661 & 99.69 & 815 & 49.48 & 1042 \\
10 & 105.97 & 958 & 15.60 & 1056 & 120.42 & 1249 & 273.00 & 1577 & 602.88 & 2273 \\
11 & 218.99 & 1582 & 28.35 & 1717 & 133.34 & 2054 & 514.07 & 2598 & 854.52 & 4008 \\
12 & 284.40 & 2442 & 700.48 & 2613 & 227.00 & 3121 & 1051.76 & 3929 & 1192.79 & 6110 \\
13 & 395.56 & 3599 & 660.53 & 3810 & 209.16 & 4486 & 1159.93 & 5605 & 1115.21 & 8592 \\
14 & 542.46 & 5095 & 809.59 & 5401 & 245.92 & 6249 & 1462.61 & 7747 & 1483.72 & 11549 \\
15 & 691.97 & 6679 & 1103.22 & 7079 & 317.13 & 8099 & 1934.87 & 10256 & 1675.93 & 14874 \\
16 & 887.42 & 8382 & 1500.57 & 8879 & 371.69 & 10074 & 2244.23 & 13269 & 1982.86 & 18703 \\
17 & 1199.76 & 10211 & 1709.38 & 10809 & 462.14 & 12174 & 2888.04 & 16399 & 2485.65 & 22656 \\
18 & 1637.79 & 12119 & 1975.14 & 12822 & 562.79 & 14356 & 3469.38 & 19558 & 2847.28 & 26634 \\
19 & 1368.93 & 14192 & 1247.71 & 14999 & 550.38 & 16699 & 3488.77 & 23078 & 2607.92 & 30971 \\
20 & 3032.49 & 16482 & 1518.02 & 17393 & 648.18 & 19263 & 3994.71 & 26680 & 3421.07 & 35397 \\
21 & 3510.55 & 18862 & 1911.90 & 19878 & 799.62 & 21924 & 4468.07 & 30462 & 4217.22 & 39998 \\
22 & 4541.46 & 21272 & 2478.25 & 22394 & 1009.26 & 24608 & 7264.24 & 34340 & 4965.61 & 44697 \\
23 & 6675.83 & 23808 & 2686.18 & 25035 & 1143.75 & 27410 & 12820.34 & 38415 & 6453.68 & 49592 \\
24 & 9760.25 & 26444 & 3368.17 & 27773 & 1446.12 & 30322 & 20498.16 & 42799 & 8577.49 & 54786 \\
25 & 18656.70 & 28936 & 3233.15 & 30368 & 1937.53 & 33088 & 28326.84 & 46977 & 8507.27 & 59784 \\
\hline
\end{tabular}
\end{sidewaystable}

\begin{sidewaystable}
    \centering
    \caption{Computing partial symmetry breaks (part 3).}
    \label{tab:appendix:part3}

\begin{tabular}{||c|||r|r||r|r||r|r||r|r||r|r||}
  \hline
  n &
    \multicolumn{2}{|c||}{ equivs(13,6,6) } &
    \multicolumn{2}{|c||}{ equivs(14,7,6)} & 
    \multicolumn{2}{|c||}{ equivs(15,7,7)}     &
    \multicolumn{2}{|c||}{ equivs(16,8,7)} & 
    \multicolumn{2}{|c||}{ equivs(17,8,8)} \\
    \hline
    &time & size &time & size &time & size &time & size &time & size   \\
  \hline
4 & 0.02    & 6 & 1.78 & 6 & 0.09 & 6 & 0.15 & 6 & 0.03 & 6 \\
5 & 0.17    & 16 & 0.04 & 16 & 0.09 & 16 & 0.09 & 16 & 0.05 & 16 \\
6 & 0.47    & 39 & 0.06 & 39 & 0.51 & 39 & 0.61 & 39 & 0.33 & 39 \\
7 & 2.49    & 117 & 0.12 & 117 & 3.51 & 117 & 4.42 & 117 & 2.10 & 117 \\
8 & 20.03   & 375 & 0.47 & 375 & 25.51 & 403 & 27.7 & 403 & 17.20 & 430 \\
9 & 138.28  & 1348 & 40.95 & 1399 & 161.30 & 1617 & 298.27 & 1621 & 162.41 & 1826 \\
10& 668.49  & 3283 & 170.23 & 3300 & 1594.38 & 5802 & 1331.38 & 6028 & 1565.23 & 7471 \\
11& 1140.05 & 6228 & 2421.84 & 6435 & 6706.86 & 13325 & 8106.02 & 14482 & 315.53 & 16877 \\
12& 1574.99 & 10000 & 3508.39 & 11004 & 10969.82 & 23352 & 1960.20 & 25422 & 598.27 & 29188 \\
13& 1769.59 & 14370 & 3219.32 & 16320 & 6643.84 & 35028 & 2892.72 & 38019 & 1126.93 & 43502 \\
14& 2334.53 & 19319 & 4190.35 & 22292 & 9124.82 & 47735 & 4089.17 & 51650 & 2273.25 & 59191 \\
15& 2284.52 & 24845 & 4833.40 & 28863 & 12007.33 & 61358 & 5238.12 & 66197 & 4802.08 & 76208 \\
16& 2901.95 & 31029 & 5104.31 & 36094 & 16943.99 & 75841 & 6711.95 & 81600 & 8697.86 & 94288 \\
17& 4750.90 & 37665 & 11393.93 & 43784 & 23268.18 & 91004 & 5553.20 & 97690 & 15954.10 & 113258 \\
18& 7772.77 & 44696 & 10034.88 & 51871 & 33153.95 & 106844 & 6748.21 & 114449 & 31299.63 & 132956 \\
19& 20943.46& 52050 & 13895.69 & 60274 & 54813.26 & 123162 & 6424.71 & 131693 & 4652.57 & 152115 \\
20& 40393.69& 59685 & 15817.37 & 68981 & 16803.40 & 137130 & 7779.78 & 146589 & 6788.29 & 169331 \\
21& 9457.51 & 66672 & 15795.77 & 77031 & 20683.14 & 150888 & 8832.33 & 161267 & 10159.39 & 186468 \\
22& 9787.25 & 73765 & 16040.81 & 85182 & 25965.64 & 164796 & 10555.06 & 176101 & 15177.3 & 203783 \\
23& 12760.56& 81142 & 16181.41 & 93624 & 30170.61 & 179009 & 14470.49 & 191243 & 22066.56 & 221458 \\
24& 16610.50& 88823 & 19426.02 & 102367 & 40860.82 & 193552 & 13608.21 & 206717 & 32226.77 & 239664 \\
25& 16306.04& 96384 & 23774.79 & 110986 & 51425.92 & 208019 & 18935.67 & 222111 & 40712.12 & 258040 \\
\hline
\end{tabular}
\end{sidewaystable}

\begin{sidewaystable}
    \centering
    \caption{Computing partial symmetry breaks (part 4).}
    \label{tab:appendix:part4}

\begin{tabular}{||c|||r|r||r|r||r|r||}
  \hline
  n &
    \multicolumn{2}{|c||}{ equivs(18,9,9) } &
    \multicolumn{2}{|c||}{ equivs(19,10,9)} & 
    \multicolumn{2}{|c||}{ equivs(20,10,10)} \\
    \hline
    &time & size &time & size &time & size    \\
  \hline
4 & 0.02 & 6 & 0.01 & 6 & 0.01 & 6 \\
5 & 0.03 & 16 & 0.04 & 16 & 0.04 & 16 \\
6 & 0.21 & 39 & 0.27 & 39 & 0.25 & 39 \\
7 & 2.07 & 117 & 2.32 & 117 & 1.98 & 117 \\
8 & 12.41 & 442 & 14.28 & 442 & 15.92 & 443 \\
9 & 110.87 & 2051 & 138.78 & 2052 & 141.86 & 2298 \\
10 & 1033.92 & 8663 & 1454.67 & 8708 & 1686.94 & 10620 \\
11 & 9854.89 & 29636 & 2044.95 & 29648 & 2255.91 & 34706 \\
12 & 3963.98 & 47777 & 3014.29 & 48559 & 3917.66 & 58897 \\
13 & 6301.31 & 68682 & 5000.8 & 70242 & 7298.59 & 86420 \\
14 & 9772.08 & 91549 & 7820.69 & 93893 & 16474.36 & 116297 \\
15 & 16033.66 & 116093 & 9300.3 & 119223 & 38950.04 & 148177 \\
16 & 24073.03 & 141837 & 11633.25 & 145755 & 93009.62 & 181601 \\
17 & 40678.21 & 168792 & 15315.38 & 173503 & 17245.82 & 214423 \\
18 & 67377.30 & 196727 & 19303.83 & 202232 & 27740.77 & 248451 \\
19 & 32050.86 & 222648 & 19169.42 & 228953 & 37616.58 & 280771 \\
20 & 39595.16 & 246903 & 21857.20 & 254014 & 60591.22 & 311677 \\
21 & 54418.02 & 271467 & 28732.12 & 279388 & 52916.92 & 341214 \\
22 & 57190.65 & 294502 & 34661.24 & 303241 & 19618.13 & 367267 \\
23 & 73856.02 & 318452 & 36254.56 & 327999 & 23073.66 & 394654 \\
24 & 47904.94 & 341114 & 43403.14 & 351562 & 25368.17 & 420852 \\
25 & 57245.46 & 364247 & 47889.13 & 375635 & 28956.97 & 447663 \\
\hline
\end{tabular}
\end{sidewaystable}

\begin{table}
  \caption{Ratios for Symmetry Breaks}\label{tab:big:ratios}
  \begin{center}
    \begin{tiny}
      \begin{tabular}[c]{|l|l|r|}
        \hline
        $n$ & Type & ratio \\
        \hline
4    & trans          & $1.00$ \\
4    & equivs(3,2,2)  & $1.00$ \\
4    & equivs(4,3,3)  & $1.00$ \\
4    & equivs(5,3,2)  & $1.00$ \\
4    & equivs(6,3,3)  & $1.00$ \\
4    & equivs(7,4,3)  & $1.00$ \\
4    & equivs(8,4,4)  & $1.00$ \\
4    & equivs(9,5,4)  & $1.00$ \\
4    & equivs(10,5,4) & $1.00$ \\
4    & equivs(11,6,5) & $1.00$ \\
4    & equivs(12,6,5) & $1.00$ \\
4    & equivs(13,6,6) & $1.00$ \\
4    & equivs(14,7,6) & $1.00$ \\
4    & equivs(15,7,7) & $1.00$ \\
4    & equivs(16,8,7) & $1.00$ \\
4    & equivs(17,8,8) & $1.00$ \\ \hline
5    & trans          & $1.26$ \\
5    & equivs(3,2,2)  & $1.00$ \\
5    & equivs(4,3,3)  & $1.00$ \\
5    & equivs(5,3,2)  & $1.00$ \\
5    & equivs(6,3,3)  & $1.00$ \\
5    & equivs(7,4,3)  & $1.00$ \\
5    & equivs(8,4,4)  & $1.00$ \\
5    & equivs(9,5,4)  & $1.00$ \\
5    & equivs(10,5,4) & $1.00$ \\
5    & equivs(11,6,5) & $1.00$ \\
5    & equivs(12,6,5) & $1.00$ \\
5    & equivs(13,6,6) & $1.00$ \\
5    & equivs(14,7,6) & $1.00$ \\
5    & equivs(15,7,7) & $1.00$ \\
5    & equivs(16,8,7) & $1.00$ \\
5    & equivs(17,8,8) & $1.00$ \\ \hline
6    & trans          & $1.77$ \\
6    & equivs(3,2,2)  & $1.24$ \\
6    & equivs(4,3,3)  & $1.08$ \\
6    & equivs(5,3,2)  & $1.00$ \\
6    & equivs(6,3,3)  & $1.00$ \\
6    & equivs(7,4,3)  & $1.00$ \\
6    & equivs(8,4,4)  & $1.00$ \\
6    & equivs(9,5,4)  & $1.00$ \\
6    & equivs(10,5,4) & $1.00$ \\
6    & equivs(11,6,5) & $1.00$ \\
6    & equivs(12,6,5) & $1.00$ \\
6    & equivs(13,6,6) & $1.00$ \\
6    & equivs(14,7,6) & $1.00$ \\
6    & equivs(15,7,7) & $1.00$ \\
6    & equivs(16,8,7) & $1.00$ \\
6    & equivs(17,8,8) & $1.00$ \\  \hline
7    & trans          & $3.02$ \\
7    & equivs(3,2,2)  & $2.00$ \\
7    & equivs(4,3,3)  & $1.54$ \\
7    & equivs(5,3,2)  & $1.39$ \\
7    & equivs(6,3,3)  & $1.15$ \\
7    & equivs(7,4,3)  & $1.07$ \\
7    & equivs(8,4,4)  & $1.02$ \\
7    & equivs(9,5,4)  & $1.02$ \\
7    & equivs(10,5,4) & $1.02$ \\
7    & equivs(11,6,5) & $1.01$ \\
7    & equivs(12,6,5) & $1.01$ \\
7    & equivs(13,6,6) & $1.00$ \\
7    & equivs(14,7,6) & $1.00$ \\
7    & equivs(15,7,7) & $1.00$ \\
7    & equivs(16,8,7) & $1.00$ \\
7    & equivs(17,8,8) & $1.00$ \\ 
         \hline
    \end{tabular}\hspace{8pt}
    \begin{tabular}[c]{|l|l|r|}
        \hline
        $n$ & Type & ratio \\
        \hline
\hline
8    & trans          & $5.39$ \\
8    & equivs(3,2,2)  & $3.55$ \\
8    & equivs(4,3,3)  & $2.47$ \\
8    & equivs(5,3,2)  & $2.24$ \\
8    & equivs(6,3,3)  & $1.70$ \\
8    & equivs(7,4,3)  & $1.59$ \\
8    & equivs(8,4,4)  & $1.23$ \\
8    & equivs(9,5,4)  & $1.13$ \\
8    & equivs(10,5,4) & $1.10$ \\
8    & equivs(11,6,5) & $1.04$ \\
8    & equivs(12,6,5) & $1.04$ \\
8    & equivs(13,6,6) & $1.02$ \\
8    & equivs(14,7,6) & $1.02$ \\
8    & equivs(15,7,7) & $1.01$ \\
8    & equivs(16,8,7) & $1.01$ \\
8    & equivs(17,8,8) & $1.00$ \\ \hline
9    & trans          & $9.42$ \\
9    & equivs(3,2,2)  & $6.32$ \\
9    & equivs(4,3,3)  & $4.08$ \\
9    & equivs(5,3,2)  & $3.76$ \\
9    & equivs(6,3,3)  & $2.82$ \\
9    & equivs(7,4,3)  & $2.61$ \\
9    & equivs(8,4,4)  & $1.68$ \\
9    & equivs(9,5,4)  & $1.49$ \\
9    & equivs(10,5,4) & $1.39$ \\
9    & equivs(11,6,5) & $1.19$ \\
9    & equivs(12,6,5) & $1.15$ \\
9    & equivs(13,6,6) & $1.11$ \\
9    & equivs(14,7,6) & $1.08$ \\
9    & equivs(15,7,7) & $1.05$ \\
9    & equivs(16,8,7) & $1.05$ \\
9    & equivs(17,8,8) & $1.03$ \\ \hline
10   & trans          & $15.34$ \\
10   & equivs(3,2,2)  & $10.61$ \\
10   & equivs(4,3,3)  & $6.54$ \\
10   & equivs(5,3,2)  & $6.10$ \\
10   & equivs(6,3,3)  & $4.61$ \\
10   & equivs(7,4,3)  & $4.21$ \\
10   & equivs(8,4,4)  & $2.35$ \\
10   & equivs(9,5,4)  & $2.04$ \\
10   & equivs(10,5,4) & $1.85$ \\
10   & equivs(11,6,5) & $1.45$ \\
10   & equivs(12,6,5) & $1.34$ \\
10   & equivs(13,6,6) & $1.23$ \\
10   & equivs(14,7,6) & $1.21$ \\
10   & equivs(15,7,7) & $1.14$ \\
10   & equivs(16,8,7) & $1.13$ \\
10   & equivs(17,8,8) & $1.07$ \\ \hline
11   & trans          & $23.52$ \\
11   & equivs(3,2,2)  & $16.81$ \\
11   & equivs(4,3,3)  & $10.21$ \\
11   & equivs(5,3,2)  & $9.53$ \\
11   & equivs(6,3,3)  & $7.23$ \\
11   & equivs(7,4,3)  & $6.51$ \\
11   & equivs(8,4,4)  & $3.25$ \\
11   & equivs(9,5,4)  & $2.78$ \\
11   & equivs(10,5,4) & $2.46$ \\
11   & equivs(11,6,5) & $1.79$ \\
11   & equivs(12,6,5) & $1.60$ \\
11   & equivs(13,6,6) & $1.42$ \\
11   & equivs(14,7,6) & $1.38$ \\
11   & equivs(15,7,7) & $1.25$ \\
11   & equivs(16,8,7) & ${\approx}1.22$ \\
11   & equivs(17,8,8) & $1.15$ \\
\hline
    \end{tabular}\hspace{8pt}
    \begin{tabular}[c]{|l|l|r|}
        \hline
        $n$ & Type & ratio \\
        \hline
 \hline
12   & trans          & ${\approx}32.47$ \\
12   & equivs(3,2,2)  & ${\approx}23.73$ \\
12   & equivs(4,3,3)  & ${\approx}14.99$ \\
12   & equivs(5,3,2)  & ${\approx}14.36$ \\
12   & equivs(6,3,3)  & ${\approx}11.24$ \\
12   & equivs(7,4,3)  & ${\approx}10.82$ \\
12   & equivs(8,4,4)  & ${\approx}4.89$ \\
12   & equivs(9,5,4)  & ${\approx}4.16$ \\
12   & equivs(10,5,4) & ${\approx}3.43$ \\
12   & equivs(11,6,5) & ${\approx}2.45$ \\
12   & equivs(12,6,5) & ${\approx}1.93$ \\
12   & equivs(13,6,6) & ${\approx}1.48$ \\
12   & equivs(14,7,6) & ${\approx}1.40$ \\
12   & equivs(15,7,7) & ${\approx}1.22$ \\
12   & equivs(16,8,7) & ${\approx}1.20$ \\
12   & equivs(17,8,8) & ${\approx}1.09$ \\ 
\hline
13   & trans          & ${\approx}48.77$ \\
13   & equivs(3,2,2)  & ${\approx}45.28$ \\
13   & equivs(4,3,3)  & ${\approx}27.87$ \\
13   & equivs(5,3,2)  & ${\approx}24.38$ \\
13   & equivs(6,3,3)  & ${\approx}16.72$ \\
13   & equivs(7,4,3)  & ${\approx}14.63$ \\
13   & equivs(8,4,4)  & ${\approx}5.57$ \\
13   & equivs(9,5,4)  & ${\approx}5.57$ \\
13   & equivs(10,5,4) & ${\approx}4.96$ \\
13   & equivs(11,6,5) & ${\approx}3.09$ \\
13   & equivs(12,6,5) & ${\approx}2.57$ \\
13   & equivs(13,6,6) & ${\approx}2.09$ \\
13   & equivs(14,7,6) & ${\approx}2.00$ \\
13   & equivs(15,7,7) & ${\approx}1.52$ \\
13   & equivs(16,8,7) & ${\approx}1.50$ \\
13   & equivs(17,8,8) & ${\approx}1.37$ \\ \hline
14   & trans          & ${\approx}69.44$ \\
14   & equivs(3,2,2)  & ${\approx}53.32$ \\
14   & equivs(4,3,3)  & ${\approx}41.54$ \\
14   & equivs(5,3,2)  & ${\approx}39.68$ \\
14   & equivs(6,3,3)  & ${\approx}26.04$ \\
14   & equivs(7,4,3)  & ${\approx}23.56$ \\
14   & equivs(8,4,4)  & ${\approx}8.37$ \\
14   & equivs(9,5,4)  & ${\approx}7.13$ \\
14   & equivs(10,5,4) & ${\approx}5.50$ \\
14   & equivs(11,6,5) & ${\approx}3.64$ \\
14   & equivs(12,6,5) & ${\approx}2.79$ \\
14   & equivs(13,6,6) & ${\approx}2.29$ \\
14   & equivs(14,7,6) & ${\approx}2.29$ \\
14   & equivs(15,7,7) & ${\approx}1.78$ \\
14   & equivs(16,8,7) & ${\approx}1.71$ \\
14   & equivs(17,8,8) & ${\approx}1.63$ \\ \hline
15   & trans          & ${\approx}71.61$ \\
15   & equivs(3,2,2)  & ${\approx}58.70$ \\
15   & equivs(4,3,3)  & ${\approx}41.09$ \\
15   & equivs(5,3,2)  & ${\approx}41.09$ \\
15   & equivs(6,3,3)  & ${\approx}33.46$ \\
15   & equivs(7,4,3)  & ${\approx}27.00$ \\
15   & equivs(8,4,4)  & ${\approx}10.27$ \\
15   & equivs(9,5,4)  & ${\approx}9.24$ \\
15   & equivs(10,5,4) & ${\approx}7.78$ \\
15   & equivs(11,6,5) & ${\approx}4.99$ \\
15   & equivs(12,6,5) & ${\approx}4.33$ \\
15   & equivs(13,6,6) & ${\approx}3.08$ \\
15   & equivs(14,7,6) & ${\approx}3.01$ \\
15   & equivs(15,7,7) & ${\approx}2.38$ \\
15   & equivs(16,8,7) & ${\approx}1.72$ \\
15   & equivs(17,8,8) & ${\approx}1.61$ \\ 
\hline
    \end{tabular}
    \end{tiny}
  \end{center}
\end{table}

\begin{table}
  \caption{Ratios for Symmetry Breaks cont.}\label{tab:big:ratios1}
  \begin{center}
    \begin{tiny}
        \begin{tabular}[c]{|l|l|r|}
        \hline
        $n$ & Type & ratio \\
        \hline
16   & trans          & ${\approx}106.25$ \\
16   & equivs(3,2,2)  & ${\approx}89.72$ \\
16   & equivs(4,3,3)  & ${\approx}67.29$ \\
16   & equivs(5,3,2)  & ${\approx}63.75$ \\
16   & equivs(6,3,3)  & ${\approx}42.50$ \\
16   & equivs(7,4,3)  & ${\approx}36.01$ \\
16   & equivs(8,4,4)  & ${\approx}14.76$ \\
16   & equivs(9,5,4)  & ${\approx}12.40$ \\
16   & equivs(10,5,4) & ${\approx}10.92$ \\
16   & equivs(11,6,5) & ${\approx}6.94$ \\
16   & equivs(12,6,5) & ${\approx}5.02$ \\
16   & equivs(13,6,6) & ${\approx}3.39$ \\
16   & equivs(14,7,6) & ${\approx}3.10$ \\
16   & equivs(15,7,7) & ${\approx}2.43$ \\
16   & equivs(16,8,7) & ${\approx}2.32$ \\
16   & equivs(17,8,8) & ${\approx}1.99$ \\ \hline
17   & trans          & ${\approx}191.28$ \\
17   & equivs(3,2,2)  & ${\approx}168.63$ \\
17   & equivs(4,3,3)  & ${\approx}123.32$ \\
17   & equivs(5,3,2)  & ${\approx}113.26$ \\
17   & equivs(6,3,3)  & ${\approx}78.02$ \\
17   & equivs(7,4,3)  & ${\approx}66.70$ \\
17   & equivs(8,4,4)  & ${\approx}22.02$ \\
17   & equivs(9,5,4)  & ${\approx}16.36$ \\
17   & equivs(10,5,4) & ${\approx}15.42$ \\
17   & equivs(11,6,5) & ${\approx}8.97$ \\
17   & equivs(12,6,5) & ${\approx}7.71$ \\
17   & equivs(13,6,6) & ${\approx}3.85$ \\
17   & equivs(14,7,6) & ${\approx}3.46$ \\
17   & equivs(15,7,7) & ${\approx}2.75$ \\
17   & equivs(16,8,7) & ${\approx}2.79$ \\
17   & equivs(17,8,8) & ${\approx}2.48$ \\ \hline
18   & trans          & ${\approx}221.25$ \\
18   & equivs(3,2,2)  & ${\approx}198.56$ \\
18   & equivs(4,3,3)  & ${\approx}127.65$ \\
18   & equivs(5,3,2)  & ${\approx}119.14$ \\
18   & equivs(6,3,3)  & ${\approx}87.93$ \\
18   & equivs(7,4,3)  & ${\approx}73.75$ \\
18   & equivs(8,4,4)  & ${\approx}34.75$ \\
18   & equivs(9,5,4)  & ${\approx}31.91$ \\
18   & equivs(10,5,4) & ${\approx}20.57$ \\
18   & equivs(11,6,5) & ${\approx}10.64$ \\
18   & equivs(12,6,5) & ${\approx}8.86$ \\
18   & equivs(13,6,6) & ${\approx}4.88$ \\
18   & equivs(14,7,6) & ${\approx}4.52$ \\
18   & equivs(15,7,7) & ${\approx}3.63$ \\
18   & equivs(16,8,7) & ${\approx}3.55$ \\
18   & equivs(17,8,8) & ${\approx}3.10$ \\  \hline
19   & trans          & ${\approx}256.27$ \\
19   & equivs(3,2,2)  & ${\approx}215.80$ \\
19   & equivs(4,3,3)  & ${\approx}151.74$ \\
19   & equivs(5,3,2)  & ${\approx}141.62$ \\
19   & equivs(6,3,3)  & ${\approx}114.65$ \\
19   & equivs(7,4,3)  & ${\approx}104.53$ \\
19   & equivs(8,4,4)  & ${\approx}43.83$ \\
19   & equivs(9,5,4)  & ${\approx}39.62$ \\
19   & equivs(10,5,4) & ${\approx}28.66$ \\
19   & equivs(11,6,5) & ${\approx}13.28$ \\
19   & equivs(12,6,5) & ${\approx}11.59$ \\
19   & equivs(13,6,6) & ${\approx}6.95$ \\
19   & equivs(14,7,6) & ${\approx}6.43$ \\
19   & equivs(15,7,7) & ${\approx}4.32$ \\
19   & equivs(16,8,7) & ${\approx}4.11$ \\
19   & equivs(17,8,8) & ${\approx}3.79$ \\ 
      \hline
    \end{tabular}\hspace{10pt}
    \begin{tabular}[c]{|l|l|r|}
        \hline
        $n$ & Type & ratio \\
      \hline
\hline
20   & trans          & ${\approx}345.82$ \\
20   & equivs(3,2,2)  & ${\approx}303.65$ \\
20   & equivs(4,3,3)  & ${\approx}227.74$ \\
20   & equivs(5,3,2)  & ${\approx}227.74$ \\
20   & equivs(6,3,3)  & ${\approx}168.69$ \\
20   & equivs(7,4,3)  & ${\approx}151.82$ \\
20   & equivs(8,4,4)  & ${\approx}60.10$ \\
20   & equivs(9,5,4)  & ${\approx}47.45$ \\
20   & equivs(10,5,4) & ${\approx}34.79$ \\
20   & equivs(11,6,5) & ${\approx}18.71$ \\
20   & equivs(12,6,5) & ${\approx}14.76$ \\
20   & equivs(13,6,6) & ${\approx}6.98$ \\
20   & equivs(14,7,6) & ${\approx}6.59$ \\
20   & equivs(15,7,7) & ${\approx}4.74$ \\ 
20   & equivs(16,8,7) & t.o. \\
20   & equivs(17,8,8) & ${\approx}3.36$ \\\hline
21   & trans          & ${\approx}465.11$ \\
21   & equivs(3,2,2)  & ${\approx}431.89$ \\
21   & equivs(4,3,3)  & ${\approx}348.84$ \\
21   & equivs(5,3,2)  & ${\approx}343.30$ \\
21   & equivs(6,3,3)  & ${\approx}215.95$ \\
21   & equivs(7,4,3)  & ${\approx}193.80$ \\
21   & equivs(8,4,4)  & ${\approx}77.52$ \\
21   & equivs(9,5,4)  & ${\approx}62.29$ \\
21   & equivs(10,5,4) & t.o. \\
21   & equivs(11,6,5) & ${\approx}22.84$ \\
21   & equivs(12,6,5) & ${\approx}17.30$ \\
21   & equivs(13,6,6) & ${\approx}8.82$ \\
21   & equivs(14,7,6) & t.o. \\
21   & equivs(15,7,7) & t.o. \\
21   & equivs(16,8,7) & t.o. \\
21   & equivs(17,8,8) & ${\approx}4.07$ \\ \hline
22   & trans          & ${\approx}639.65$ \\
22   & equivs(3,2,2)  & ${\approx}624.42$ \\
22   & equivs(4,3,3)  & ${\approx}380.74$ \\
22   & equivs(5,3,2)  & ${\approx}380.74$ \\
22   & equivs(6,3,3)  & ${\approx}281.75$ \\
22   & equivs(7,4,3)  & ${\approx}247.48$ \\
22   & equivs(8,4,4)  & ${\approx}102.80$ \\
22   & equivs(9,5,4)  & ${\approx}79.96$ \\
22   & equivs(10,5,4) & ${\approx}68.53$ \\ 
22   & equivs(11,6,5) & t.o. \\
22   & equivs(12,6,5) & t.o. \\
22   & equivs(13,6,6) & t.o. \\
22   & equivs(14,7,6) & t.o. \\
22   & equivs(15,7,7) & t.o. \\
22   & equivs(16,8,7) & t.o. \\
22   & equivs(17,8,8) & t.o. \\ \hline
23   & trans          & t.o. \\
23   & equivs(3,2,2)  & t.o. \\
23   & equivs(4,3,3)  & t.o. \\     
23   & equivs(5,3,2)  & ${\approx}591.16$ \\
23   & equivs(6,3,3)  & ${\approx}470.74$ \\
23   & equivs(7,4,3)  & ${\approx}383.16$ \\ 
23   & equivs(8,4,4)   & t.o. \\
23   & equivs(9,5,4)   & t.o. \\
23   & equivs(10,5,4)  & t.o. \\
23   & equivs(11,6,5)  & t.o. \\
23   & equivs(12,6,5)  & t.o. \\
23   & equivs(13,6,6)  & t.o. \\
23   & equivs(14,7,6)  & t.o. \\
23   & equivs(15,7,7)  & t.o. \\
23   & equivs(16,8,7)  & t.o. \\
23   & equivs(17,8,8)  & t.o. \\
      \hline
    \end{tabular}
  \end{tiny}
   \end{center}
 \end{table}

\end{document}